\documentclass[
	groupedaddress,
	amsmath,amssymb,
	prb,
	floatfix,
	superscriptaddress,
    twocolumn,
	aps
 ]{revtex4-2}

\usepackage{graphicx}
\usepackage{dcolumn}
\usepackage{bm}
\usepackage[mathlines]{lineno}
\usepackage{xcolor}
\usepackage{soul}
\usepackage{adjustbox}
\usepackage{nth}
\usepackage{hyperref}
\usepackage{cleveref}
\usepackage[T1]{fontenc}
\usepackage{nth}
\usepackage{amsmath}
\usepackage{amssymb}
\usepackage{xr}

\newcommand{\sicref}[1]{\cref{#1} in the SI}

\begin{document}

\title{Diffusion coefficients of multi-principal element alloys from first-principles} 
\author{Damien K. J. Lee}
\affiliation{Laboratory of materials design and simulation (MADES), Institute of Materials, \'{E}cole Polytechnique F\'{e}d\'{e}rale de Lausanne}
\affiliation{National Centre for Computational Design and Discovery of Novel Materials (MARVEL), \'{E}cole Polytechnique F\'{e}d\'{e}rale de Lausanne}
\author{Anirudh Raju Natarajan}
\email{anirudh.natarajan@epfl.ch}
\affiliation{Laboratory of materials design and simulation (MADES), Institute of Materials, \'{E}cole Polytechnique F\'{e}d\'{e}rale de Lausanne}
\affiliation{National Centre for Computational Design and Discovery of Novel Materials (MARVEL), \'{E}cole Polytechnique F\'{e}d\'{e}rale de Lausanne}

\begin{abstract}
  Vacancy-mediated diffusion in multi-principal element alloys (MPEAs) remains poorly understood. Existing computational methods face challenges in connecting electronic structure to macroscopic transport coefficients due to the large number of chemical elements. To address this, we introduce the embedded local cluster expansion (eLCE), which bridges first-principles calculations with kinetic Monte Carlo simulations to compute the matrix of multicomponent diffusion coefficients. Applying this approach to refractory MPEAs in the V--Cr--Nb--Mo--Ta--W system, we evaluate the complete mobility and diffusion tensors of a six-component alloy at finite temperatures. We find that, in these refractory alloys, local kinetic barriers, rather than thermodynamics or vacancy correlation factors, primarily control diffusion. Whether vacancy diffusion is sluggish or anti-sluggish depends on the mean vacancy migration barrier relative to the rule-of-mixtures estimate and on the availability of percolating pathways of fast-diffusing species. We use this insight to screen the senary composition space and identify compositions with anti-sluggish vacancy diffusion. This study presents a predictive, first-principles approach for computing non-dilute transport coefficients and designing MPEAs with targeted transport properties, and more broadly establishes a route to modeling vacancy-mediated diffusion in compositionally complex crystalline materials.
\end{abstract}

\maketitle

\section{Introduction}
Diffusion governs microstructural evolution, phase stability, creep resistance, and high-temperature performance in alloys \cite{ardell_trans-interface_2005,miracle_critical_2017,darling_extreme_2016,hinrichs_ductile_2025,wang_multiscale_2025}. In functional materials such as batteries, catalysts, and thermoelectrics, diffusion governs transport, reaction kinetics, and overall device performance \cite{chen_unlocking_2024,jun_diffusion_2024,chmelik_situ_2010,panda_diffusion-programmed_2025,dahan_first-principles_2021,zhang_direct_2021}. Multi-principal element alloys (MPEAs) are of growing interest for high-temperature applications \cite{yeh_nanostructured_2004,miracle_critical_2017}, yet diffusion in these systems remains poorly understood. Diffusion in MPEAs has been widely described as ``sluggish'' \cite{tsai_sluggish_2013,hsu_clarifying_2024}, but the microscopic origin of this claim has not been established, largely because a quantitative, first-principles evaluation of diffusion in these alloys has not been achieved.

Quantifying diffusion in multicomponent alloys is challenging from both experimental and computational perspectives. Experimentally, coupled fluxes and cross-correlations between species complicate the extraction of transport coefficients \cite{paul_thermodynamics_2014,dehoff_trouble_2002,dash_recent_2022}. Computationally, modeling vacancy-mediated diffusion requires resolving the combinatorial complexity of local chemical environments that govern migration barriers \cite{xing_neural_2024,goiri_role_2019}. Furthermore, simulations must access timescales long enough to yield macroscopic transport coefficients. Most existing studies rely on tracer diffusivities \cite{zhang_zr_2022,sen_anti-sluggish_2023} or pseudo-binary approximations \cite{kirkaldy_diffusion_1966,esakkiraja_pseudo-binary_2019,dash_solving_2020}. A rigorous evaluation of the complete multicomponent diffusion tensor from first-principles statistical mechanics has not been achieved beyond simple alloys \cite{dash_recent_2022,van_der_ven_rechargeable_2020}.

At the atomistic scale, diffusion in substitutional alloys proceeds through thermally activated vacancy hops \cite{van_der_ven_vacancy_2010,balluffi_kinetics_2005,kirkaldy_diffusion_1987,shewmon_thermal_1958}. In multicomponent systems, the migration barrier for each hop depends on the local chemical environment, producing a broad spectrum of activation energies even in binary alloys \cite{goiri_role_2019,behara_role_2024}. A rigorous description of diffusion therefore requires both an accurate model of migration energies across diverse local environments and sufficient statistical sampling of vacancy transport over long timescales. Cluster expansion techniques coupled with kinetic Monte Carlo (kMC) simulations can in principle provide both ingredients \cite{deng_fundamental_2022,jadidi_kinetics_2023,canepa_pushing_2023}. Although cluster expansions have been highly successful for modeling configurational thermodynamics and vacancy migration in simpler alloys, their extension to kinetic barriers in concentrated multicomponent systems remains computationally demanding and has long been considered intractable \cite{dash_recent_2022,xing_neural_2024}.

Here, we introduce the embedded local cluster expansion (eLCE) formalism, which enables rapid and accurate evaluation of kinetically resolved activation barriers across complex multicomponent environments. By coupling this formalism with kMC simulations and first-principles electronic structure calculations, we study substitutional diffusion in the senary V--Cr--Nb--Mo--Ta--W system, a refractory MPEA under active investigation for its mechanical and high-temperature properties \cite{cook_kink_2024,kumar_degradation_2024,george_high-entropy_2019,mak_ductility_2021,li_complex_2020}. We rigorously compute the full non-dilute diffusion coefficient tensor from first-principles statistical mechanics and analyze diffusion in terms of its collective eigenmodes. Our results reveal that sluggish diffusion in this class of alloys is governed by the structure of the vacancy barrier spectrum and the emergence of percolating low-energy pathways, rather than chemical complexity alone. This work establishes a predictive framework for diffusion in multicomponent alloys, rigorously connecting atomistic migration energies to macroscopic transport coefficients. More broadly, the framework is general and can be extended to other crystalline systems in which diffusion controls functional properties.

\section{Results}
Our goal is to compute the diffusion coefficients of a multicomponent alloy directly from its electronic structure. Diffusion in these materials proceeds through a sequence of atomic hops, in which an atom exchanges position with a neighboring vacancy. Each hop must overcome an energy barrier, and the rate of diffusion is set by how often these barriers are crossed. The difficulty specific to multicomponent alloys is that the barrier for a given hop depends on the arrangement of atoms surrounding it, and the number of distinct local arrangements grows combinatorially with the number of elements. Capturing this dependence across the full range of environments has been the central obstacle to first-principles diffusion modeling in these systems. Our approach separates the problem into three stages, summarized in \cref{fig:kra}c. First, we use first-principles calculations to compute migration barriers for a set of representative local environments. Second, we train a compact model, the embedded local cluster expansion (eLCE), that predicts the barrier for any local arrangement from these data, sidestepping the need to calculate every environment explicitly. Third, we use the trained model to drive kinetic Monte Carlo simulations, which follow vacancy motion over long timescales and yield the macroscopic transport coefficients. The sections below develop each stage in turn, beginning with the formalism that connects atomic hops to the measured diffusion coefficients.

\subsection{Kinetics of multicomponent alloys}
We begin by outlining the formalism required to compute kinetic properties from first principles. In a perfect crystal, the number of lattice sites, including vacancies, is conserved. Diffusion is driven by gradients in diffusion potentials \cite{kehr_mobility_1989,balluffi_kinetics_2005}, defined as the difference between the chemical potential of species $i$ and that of vacancies:
\begin{equation}
    \tilde{\mu}_i=\mu_i-\mu_{Va}
\end{equation}
The diffusion fluxes of the $c$ chemical elements in the multicomponent alloy are related to the diffusion potentials through the Onsager transport coefficients:
\begin{equation}
\begin{bmatrix}
\vec{J_1} \\
\vec{J_2} \\
\vdots \\
\vec{J_c}
\end{bmatrix}=
-\begin{bmatrix}
L_{11} & L_{12} & \cdots & L_{1c} \\
L_{21} & L_{22} & \cdots & L_{2c} \\
\vdots & \vdots & \ddots & \vdots \\
L_{c1} & L_{c2} & \cdots & L_{cc}
\end{bmatrix}
\begin{bmatrix}
\nabla \tilde{\mu}_1 \\
\nabla \tilde{\mu}_2 \\
\vdots \\
\nabla \tilde{\mu}_c
\end{bmatrix}
\label{eq:flux}
\end{equation}
where $L_{ij}$ are the elements of the Onsager transport coefficient matrix $\mathbf{L}$. Since the total number of sites is conserved, only the fluxes of migrating species need to be tracked, with $\sum_i\vec{J_i} +\vec{J}_{Va}=0$.

The Onsager transport coefficients can be obtained using Kubo-Green linear response theory \cite{green_markoff_1954,kubo_statistical-mechanical_1957}:

\begin{align}
L_{ij} &= \frac{1}{\Omega k_B T} \tilde{L}_{ij} \\
\tilde{L}_{ij} &=
\frac{1}{2d\, t\, M}
\left\langle
\left( \sum_\zeta \Delta \vec{R}_i^\zeta(t) \right) \cdot
\left( \sum_\zeta \Delta \vec{R}_j^\zeta(t) \right)
\right\rangle
\label{eq:kubo_green}
\end{align}
in which $\Omega$ is the volume per substitutional site, $d$ is the dimensionality of the diffusion network, and $M$ is the number of substitutional sites in the crystal. The vectors $\Delta\vec{R}_i^\zeta(t)$ connect the endpoints of the trajectory of atom $\zeta$ of species $i$ after time $t$. These trajectories can be obtained from simulations in which atoms exchange with vacancies at frequencies given by transition state theory \cite{vineyard_frequency_1957}:
\begin{equation}
    \Gamma=\nu^*\exp(\frac{-\Delta E}{k_BT})
\label{eqn:transition_state}
\end{equation}
where $\nu^*$ is a vibrational prefactor and $\Delta E$ is the migration barrier of a hop.
When $\Delta E$ is known for each possible hop, kMC simulations can be used to evolve the system and compute the transport coefficients from \cref{eq:kubo_green} \cite{binder_monte_2010}.

Gradients in chemical potentials are not readily accessible in experiments. Instead, the diffusion coefficient matrix $\mathbf{D}$, which relates fluxes to concentration gradients, is more commonly used. The non-dilute diffusion coefficients can be derived from the Onsager transport coefficients \cite{balluffi_kinetics_2005,van_der_ven_vacancy_2010,kirkaldy_diffusion_1987,shewmon_thermal_1958}:

\begin{equation}
\mathbf{D} = 
\begin{bmatrix}
D_{11} & D_{12} & \cdots & D_{1c} \\
D_{21} & D_{22} & \cdots & D_{2c} \\
\vdots & \vdots & \ddots & \vdots \\
D_{c1} & D_{c2} & \cdots & D_{cc}
\end{bmatrix}=
    \mathbf{\tilde{L}} \mathbf{\tilde{\Theta}}
\end{equation}

where the thermodynamic factor matrix $\boldsymbol{\tilde{\Theta}}$ is proportional to the Hessian of the free energy, and can be calculated from composition fluctuations within the semi-grand canonical ensemble:
\begin{equation}
    \tilde{\Theta}^{-1}_{ij}=\frac{1}{M}\langle N_i N_j \rangle - \langle N_i \rangle \langle N_j \rangle
\end{equation}
Both $\mathbf{\tilde{L}}$ and $\mathbf{\tilde{\Theta}}$ are computed under boundary conditions of fixed overall composition, vacancy concentration, and temperature.

Additional metrics characterize different aspects of diffusion. The correlation factor $f_i$ measures the degree of correlation between successive hops:
\begin{equation}
    f_i=\frac{\sum_\zeta \Delta \vec{R}_i^\zeta(t)^2}{\sum_\zeta N_i^\zeta(t) \Delta\vec{r}^2}
\end{equation}
where $N^\zeta(t)$ is the number of hops performed by atom $\zeta$ and $\Delta\vec{r}^2$ is the square of the elementary hop distance. The tracer diffusion coefficient measures the diffusivity of tagged atoms in the absence of chemical gradients:
\begin{equation}
    D^*_i=\frac{\langle \Delta \vec{R}_i^2 \rangle}{2dt}
\end{equation}
Tracer diffusion coefficients can be obtained experimentally using the radiotracer method and are one of the most commonly reported quantity in diffusion studies.

\subsection{Local Cluster Expansion Formalism}
Traditional cluster expansions \cite{sanchez_generalized_1984} have primarily been applied to global properties, such as the total energy of a crystal, and have proven effective in capturing how these properties depend on the chemical decoration. However, several local properties, such as the magnetic moment \cite{ozolins_notitle_1996}, local vibrational entropy \cite{morgan_local_1998}, and generalized stacking fault energies \cite{natarajan_linking_2020}, also exhibit dependence on the chemical arrangement. These properties can be modeled using a \textit{local cluster expansion} (LCE). The primary distinction between global and local cluster expansions lies in the symmetry operations used to simplify the expansion. Global expansions exploit symmetries of the entire crystal, whereas local expansions focus on the symmetry operations that leave the local quantity invariant. 

\begin{figure*}
    \centering 
    \includegraphics{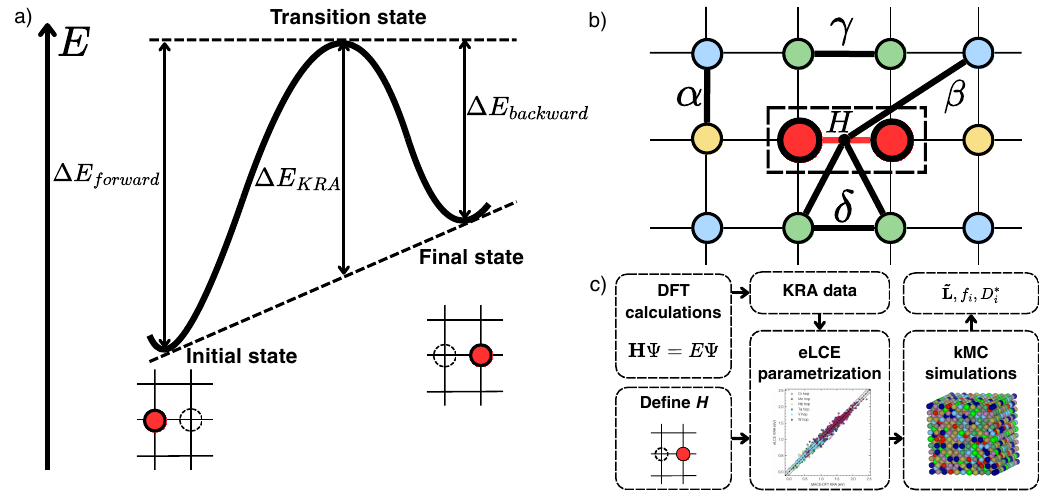}
    \caption{\textbf{a)} Energy landscape for a vacancy hop between initial and final states. The kinetically resolved activation (KRA) barrier, $\Delta E_{KRA}$, isolates the local contribution to the migration energy by subtracting the average end-state energy from the transition-state energy. The red circle denotes the hopping atom and the dashed circle denotes the vacancy. \textbf{b)} Local cluster expansion on a square lattice. The hop cluster $H$ (red) defines the migrating atom--vacancy pair. Symmetrically equivalent clusters in the surrounding environment are colored identically. \textbf{c)} Computational workflow. DFT calculations provide KRA training data for the eLCE parametrization, which enables rapid evaluation of migration barriers across arbitrary local environments. The parameterized eLCE model is then used in kMC simulations to compute transport coefficients ($\mathbf{\tilde{L}}$, $f_i$, $D^*_i$).}
    \label{fig:kra}
\end{figure*}

A migration event involves a single atom exchanging places with a vacancy. The migration barrier $\Delta E$ (\cref{eqn:transition_state}) depends on the local chemical ordering surrounding the hop. In multicomponent alloys, the barrier also depends on the direction of the hop. This directional dependence is illustrated in \cref{fig:kra}a, where $\Delta E_{forward}$ differs from $\Delta E_{backward}$.

To decouple the local and long-range energy contributions, the kinetically resolved activation barrier (KRA) was introduced by \citeauthor{van_der_ven_first-principles_2001} \cite{van_der_ven_first-principles_2001}. The KRA is defined as the difference between the transition-state energy and the average of the two end-state energies:
\begin{equation}
    \Delta E_{KRA} = E_{\mathrm{transition}}-\frac{(E_{\mathrm{initial}}+E_{\mathrm{final}})}{2}
\label{eqn:kra}
\end{equation}
which is illustrated in \cref{fig:kra}a. We define the hop cluster $H$ as the pair cluster consisting of the vacancy and the migrating atom. The occupational variable $\sigma_H$ encodes the identity of the atom performing the hop. $\Delta E_{\text{KRA}}$ therefore depends on $\sigma_H$, as well as the local ordering vector $\vec{\sigma}_L=[\sigma_1,\sigma_2,\cdots,\sigma_N]$, which collects the occupational variables of all $N$ lattice sites in the crystal excluding $H$. In practice, the LCE is truncated and depends only on a subset of sites in the immediate vicinity of the hop cluster. The end-state energies, $E_{\mathrm{initial}}$ and $E_{\mathrm{final}}$, can typically be obtained from an energy cluster expansion, allowing the forward and backward migration barriers to be recovered from the KRA.


The configurational dependence of the KRA barrier, $\Delta E_{\text{KRA}}(\vec{\sigma}_L,\sigma_H)$, is expanded with a local cluster expansion as \cite{van_der_ven_first-principles_2001}:
\begin{equation}
    \Delta E_{KRA}(\vec{\sigma}_L,\sigma_H)=K_0+\sum_{\alpha,\beta} K_{\alpha\beta}\Phi_{\alpha\beta}(\vec{\sigma}_L,\sigma_H)
\label{eqn:cluster_expansion}
\end{equation}
where $K_{\alpha\beta}$ are expansion coefficients determined during training and $\Phi_{\alpha\beta}$ are cluster functions defined as:
\begin{equation}
\label{eqn:cluster_functions}
    \Phi_{\alpha\beta}=
    \psi_{\beta}(\sigma_H)\prod_{(i,\nu)\in\alpha}\phi_{\nu}(\sigma_i)
\end{equation}
Here, $i$ is the site index, $\phi_{\nu}$ is the site basis function with index $\nu$, and $\psi_{\beta}$ is the site basis function with index $\beta$ associated with $H$. As with traditional cluster expansions, the local cluster expansion can be interpreted as interactions arising from points, pairs, triplets, and higher-order terms. As shown in \sicref{SI:sec:LCE_decomposition}, using the occupational basis as the site basis functions for the hop cluster is functionally equivalent to constructing separate models for each hopping species, which has been the approach adopted in previous studies \cite{goiri_role_2019,behara_role_2024,van_der_ven_vacancy_2010,zhang_ab_2022}.

Symmetry reduces the number of independent terms in the expansion of \cref{eqn:cluster_expansion}. In a local cluster expansion, symmetry equivalence is determined by a subgroup $\mathcal{H} \subseteq \mathcal{S}$ (where $\mathcal{S}$ is the space group of the crystal) consisting of point group operations that leave the hop cluster invariant. All symmetrically equivalent cluster functions can be grouped into an \textit{orbit} and share the same expansion coefficient. Once all symmetrically distinct orbits have been identified, the LCE can be parametrized using first-principles data. Importantly, since all symmetry operations in the local cluster expansion are defined relative to the center of $H$, the inclusion of $H$ in cluster functions does not break symmetry. As a result, all symmetrically distinct orbits in the local environment can directly incorporate descriptors of the hop cluster. This symmetry grouping is illustrated in an undecorated 2D square lattice in \cref{fig:kra}b, where point clusters in the local environment are colored by symmetry equivalence. An example of how symmetry reduces the number of independent terms is given in \sicref{SI:sec:symmetry}. The cluster labeled $\gamma$ (excluding the hopping atom) becomes cluster $\delta$, and the point cluster $\beta$ transforms to a pair cluster upon including $H$. Symmetrically-equivalent point clusters in the nearest-neighbor sites of the hop cluster in the bcc lattice are displayed in \sicref{SI:fig:bcc_symmetry}.

\subsection{Embedded Local Cluster Expansions}
A $c$-component alloy requires $c$ site basis functions to describe all possible occupants on a site. The values of these basis functions can be collected in a $c\times c$ matrix:
\begin{equation}
\mathbf{\Phi}=
\begin{bmatrix}
    \phi_1(1) & \phi_1(2) & \cdots & \phi_1(c) \\
    \phi_2(1) & \phi_2(2) & \cdots & \phi_2(c) \\
    \vdots & \vdots & \ddots & \vdots \\
    \phi_c(1) & \phi_c(2) & \cdots & \phi_c(c)
\end{bmatrix}
\end{equation}
where column $i$ of $\Phi$ collects the values of the $c$ site basis functions when the $i^{\textrm{th}}$ element occupies the site.

As with traditional cluster expansions, the number of cluster functions in the LCE increases polynomially with the number of species \cite{barroso-luque_cluster_2024,muller_constructing_2025}. To address this curse of dimensionality, we follow the embedded cluster expansion (eCE) formalism of \citeauthor{muller_constructing_2025} \cite{muller_constructing_2025} and embed the site basis functions in a lower-dimensional space through a linear transformation:

\begin{equation}
    \label{eq:eCE_site_basis_functions}
    \vec{\tilde{\phi}}(\sigma_i) = T\vec{\phi}(\sigma_i)
\end{equation}

Here, $T$ is a $k \times c$ learnable embedding matrix ($k<c$) and $\vec{\tilde{\phi}}(\sigma_i)$ denotes the embedded site basis functions. The embedding reduces the number of basis functions per site, keeping the resulting cluster expansion computationally tractable.

Local descriptors of chemical ordering are constructed by taking tensor products of the embedded basis functions across sites within a cutoff radius around the hop cluster. These tensor products are symmetrized according to the symmetry of the hop cluster and crystal, guaranteeing invariance under symmetry operations. The resulting symmetrized descriptors serve as inputs to linear or nonlinear models that predict the KRA barrier. eCE models have been shown to achieve comparable accuracy to traditional cluster expansions while requiring substantially fewer training data, by leveraging chemical similarities among elements to reduce dimensionality \cite{muller_constructing_2025}.

In this work, we extend the eCE formalism to local cluster expansions, yielding the eLCE. Rapid and accurate barrier predictions are essential for large-scale kMC simulations that access timescales well beyond the reach of molecular dynamics. We demonstrate the eLCE in the senary V--Cr--Nb--Mo--Ta--W system, with the overall workflow summarized in \cref{fig:kra}c.

\subsection{Construction of eLCE models}

\begin{figure}[h!]
    \centering
    \includegraphics{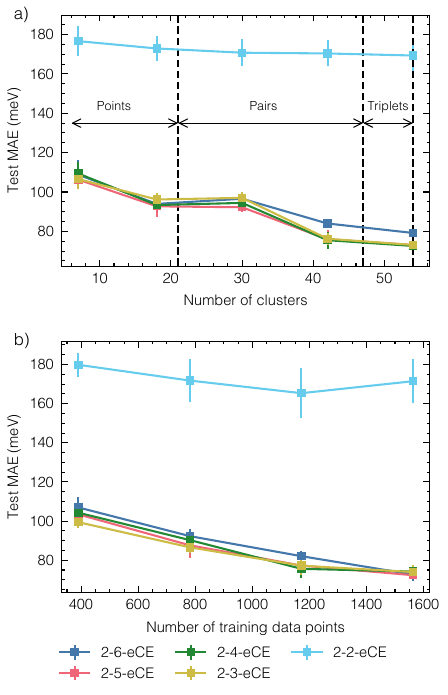}
    \caption{\textbf{a)} Test mean absolute error (MAE) of the eLCE models as a function of the number of clusters, grouped by cluster type (points, pairs, triplets). Dashed vertical lines separate the cluster types. \textbf{b)} Test MAE as a function of the number of training data points. In both panels, each curve corresponds to a different embedding dimension for the local environment ($m = 2$--6), with the hop cluster embedding fixed at $n = 2$. Error bars indicate the standard deviation over five independent random initializations of the training set and neural network weights.}
    \label{fig:learning_curves}
\end{figure}

Parameterizing eLCE models requires accurate first-principles training data across the vast compositional space of MPEAs. To make this tractable, we leverage the MACE-MPA0 foundation model \cite{batatia_mace_2023} to relax end-state and transition-state configurations, after which single-point density functional theory (DFT) calculations are performed to obtain the KRA barriers. This workflow achieves \textit{ab initio}-level accuracy at a fraction of the computational cost, as validated against full DFT nudged-elastic-band calculations (\sicref{SI:sec:mace_dft_comparison}). Further details on the workflow and dataset generation are provided in \cref{sec:methods}.

\Cref{fig:learning_curves} shows the learning curves of our eLCE models, trained on 1,945 data points with an 8:1:1 training/validation/test split. At each point on the learning curve, five models were trained from independent random initializations of the training dataset and neural network weights. Two separate embedding matrices were learned for the hop cluster and the local environment. In all models, the embedding dimension for the hop cluster was fixed at $n=2$, while the embedding dimension for neighboring sites, $m$, was systematically varied. Each model is denoted $n$-$m$-eLCE, where $n$ and $m$ correspond to the embedding dimensions of the hop cluster and local environment, respectively. Varying the hop cluster embedding dimension showed a negligible effect on model accuracy. Further details on the training procedure are provided in \cref{sec:methods}.

Embedding dimensions of $m = 3$--6 consistently yield the best performance, achieving validation mean absolute errors (MAE) of approximately 75~meV with roughly 55 clusters and 1,200 data points. Consistent with prior observations \cite{muller_constructing_2025}, an embedding dimension of $m=2$ performs poorly, with errors of $\sim$180~meV even with the full set of clusters and data.

The final selected model, a 2-3-eLCE, achieves a MAE of 69.9~meV on the unbiased test set. A parity plot comparing eLCE predictions to DFT-calculated barriers for each hopping species is shown in \sicref{SI:fig:parity_plot_eLCE}. A separate 3-eCE model was trained to reproduce the configurational energies of the group 5--6 elements and vacancies on the body-centered cubic (bcc) lattice. This energy model achieves a MAE of 3.5~meV/site and accurately reproduces the end-state energies for each hop.

\subsection{Kinetic Monte Carlo simulations}

\begin{figure}[h!]
    \centering
    \includegraphics{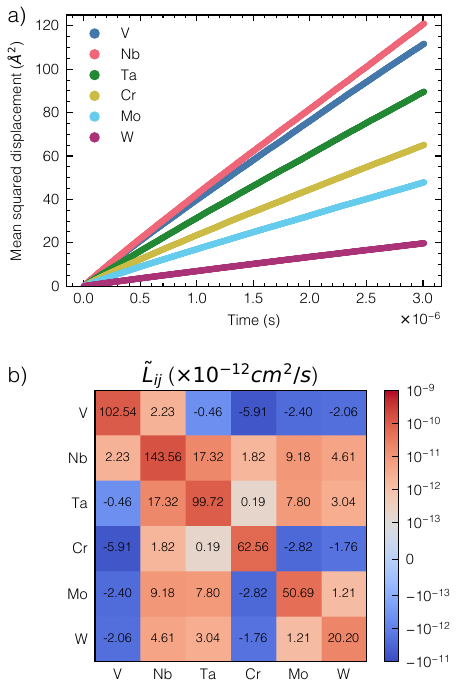}
    \caption{\textbf{a)} Mean squared displacement of each element in equiatomic VCrNbMoTaW as a function of kMC simulation time, showing that group 5 elements (V, Nb, Ta) diffuse faster than group 6 elements (Cr, Mo, W). \textbf{b)} Onsager transport coefficient matrix $\mathbf{\tilde{L}}$ in the equiatomic VCrNbMoTaW alloy. Both panels are computed at 2000\,K. All values indicated in the matrix are reported in units of $10^{-12}$ cm$^2$/s. For example, the value of $\tilde{L}_{W,W}$ is $20.20\times10^{-12}$ cm$^2$/s.}
    \label{fig:L_matrix}
\end{figure}

We demonstrate the eLCE framework through kMC simulations of vacancy diffusion in equiatomic VCrNbMoTaW, which forms a disordered solid solution on the bcc structure at the temperatures of interest \cite{miracle_critical_2017,miracle_strength_2024,natarajan_crystallography_2020}.

We computed $\mathbf{\tilde{L}}$, $f_i$, and $D^*_i$ across a range of temperatures using 500 independent kMC trajectories of 100,000 steps each on a 2000-atom simulation cell. To capture thermodynamically relevant short-range order (SRO), initial configurations were generated using canonical Monte Carlo at each temperature. In the dilute-vacancy limit, $\mathbf{\tilde{L}}$ and $D_{i\neq Va}^*$ scale linearly with vacancy concentration \cite{van_der_ven_vacancy_2010} and were rescaled using the equilibrium vacancy concentration obtained from a previously developed coarse-graining method \cite{lee_modeling_2026}. The thermodynamic factor matrix $\mathbf{\tilde{\Theta}}$ was also calculated with a recently developed coarse-graining method \cite{lee_first-principles_2026}. Using semi-grand canonical Monte Carlo simulations, the order-disorder temperature for this alloy was determined to be approximately 1900 K \cite{lee_modeling_2026}. Calculations were therefore performed at 2000 K and above to ensure the alloy remains in the single-phase disordered state. The workflow can in principle be applied at lower temperatures to study transport in any phase that is thermodynamically stable over the temperature range of interest.

\Cref{fig:L_matrix} presents results at 2000~K. The mean squared displacement (MSD) in \cref{fig:L_matrix}a extends to timescales on the order of 1~$\mu$s, orders of magnitude beyond those attainable with conventional molecular dynamics. The mobilities follow the order Nb$>$V$>$Ta$>$Cr$>$Mo$>$W, indicating that group 5 elements diffuse more rapidly than group 6 elements.

\Cref{fig:L_matrix}b shows the matrix of Onsager transport coefficients computed at 2000 K. The matrix highlights the relative magnitudes of the diagonal (self-transport) and off-diagonal (cross-correlation) terms. The diagonal terms are related to the tracer diffusion coefficients \cite{darken_formal_1951}, and therefore follow the same trends observed in the MSD of \cref{fig:L_matrix}a. The off-diagonal terms reflect the strength of kinetic coupling between different species. For example, $L_{NbTa}$ has the largest magnitude among the cross terms, indicating that a chemical potential gradient in Nb can drive Ta diffusion and vice versa. This likely arises from the chemical similarity between these two elements.

\begin{figure*}
    \centering
    \includegraphics{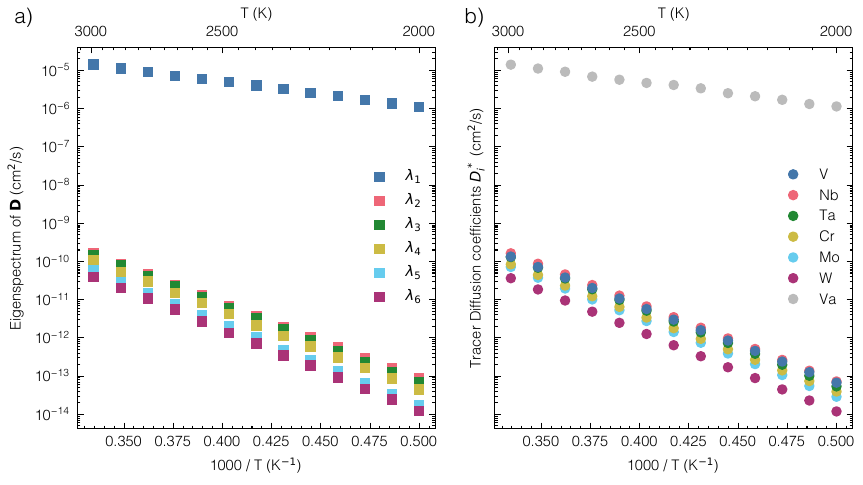}
    \caption{\textbf{a)} Eigenvalues of the diffusion coefficient matrix $\mathbf{D}$ in equiatomic VCrNbMoTaW as a function of inverse temperature. The largest eigenvalue $\lambda_1$ corresponds to the vacancy tracer diffusion coefficient and is separated from the remaining five eigenvalues ($\lambda_{2}$--$\lambda_{6}$) by several orders of magnitude. \textbf{b)} Tracer diffusion coefficients $D^*_i$ of each element and the vacancy as a function of inverse temperature. The vacancy tracer diffusivity matches $\lambda_1$, while the elemental tracer diffusivities are of the same order as $\lambda_{2}$--$\lambda_{6}$.}
    \label{fig:eigenspectrum}
\end{figure*}

From $\mathbf{\tilde{L}}$ and $\mathbf{\tilde{\Theta}}$, we computed the diffusion coefficient matrix $\mathbf{D}$. Because the individual entries of $\mathbf{D}$ are difficult to interpret directly, we diagonalize $\mathbf{D}$ to extract physically meaningful transport modes. In a binary alloy, \citeauthor{kehr_mobility_1989} showed that the larger eigenvalue of $\mathbf{D}$ corresponds to the vacancy tracer diffusion coefficient, while the smaller eigenvalue represents an interdiffusion mode in the reference frame of the crystal \cite{kehr_mobility_1989}. We computed the eigenspectrum of $\mathbf{D}$ as a function of temperature (\cref{fig:eigenspectrum}a) and compared it with the tracer diffusion coefficients (\cref{fig:eigenspectrum}b).

Consistent with the analysis of \citeauthor{kehr_mobility_1989}, the largest eigenvalue $\lambda_1$ closely matches the vacancy tracer diffusion coefficient, while the remaining eigenvalues $\lambda_{2\text{--}6}$ are of the same order as the elemental tracer diffusion coefficients. After normalizing $\lambda_{2\text{--}6}$ by the vacancy concentration, $\lambda_{2\text{--}4}$ have magnitudes comparable to $\lambda_1$, whereas $\lambda_{5\text{--}6}$ are smaller by a factor of approximately 3--4. As in binary alloys \cite{van_der_ven_vacancy_2010}, the smaller eigenvalues likely scale to first order with vacancy concentration and can be interpreted as distinct interdiffusion modes in the crystal frame of reference. These results indicate that the fastest diffusion mode arises from vacancy transport, while the slower interdiffusion modes are primarily limited by the low vacancy concentration in the alloy.

\section{Discussion}
The eLCE formalism presented in this study addresses the long-standing challenge of rigorously computing non-dilute transport coefficients in multicomponent alloys from first principles. Using this framework, we evaluate the full diffusion tensor $\mathbf{D}$ in an MPEA while explicitly accounting for the combinatorial complexity of local chemical environments and finite-temperature effects. To the best of our knowledge, this represents the first quantitative evaluation of non-dilute diffusion coefficients in a compositionally complex alloy at finite temperature from first principles. Previous attempts to extend conventional cluster expansions to migration barriers in concentrated alloys have faced two major obstacles, namely the high computational cost of generating sufficiently diverse first-principles training data and the rapid growth in fitting parameters with the number of components \cite{zhang_ab_2022,xing_neural_2024}. The eLCE framework overcomes these limitations by introducing a low-dimensional embedding that exploits chemical similarity between species, significantly reducing the number of basis functions and fitting parameters while retaining near first-principles accuracy. The resulting model enables large-scale kinetic Monte Carlo simulations with well-converged statistics at experimentally relevant timescales, providing a rigorous and computationally tractable pathway for predictive diffusion modeling in compositionally complex alloys.

One of the most enduring debates in the field of MPEAs is whether increasing the number of alloy components leads to sluggish diffusion \cite{yeh_nanostructured_2004,yeh_recent_2006,yeh_high-entropy_2007,tsai_sluggish_2013,miracle_high-entropy_2017}. The concept was introduced by \citeauthor{tsai_sluggish_2013} \cite{tsai_sluggish_2013}, who compared elemental tracer diffusion coefficients in CoCrFeMnNi to those in reference alloys and attributed the observed reduction in mobility to variations in site potential energies that create ``vacancy traps'' and to a distribution of saddle-point energies in which high barriers impede diffusion. This study sparked wide-ranging debate around the presence or absence of sluggish diffusion and the validity of simple rule-of-mixtures (ROM) estimates for transport coefficients in multicomponent alloys. However, these debates have been difficult to resolve without quantitatively reliable diffusion models for compositionally complex systems. Having established such a framework, we now systematically examine the factors that contribute to the existence or absence of sluggish diffusion in refractory alloys composed of elements in groups 5 and 6 of the periodic table.

Establishing the presence or absence of sluggish diffusion requires both a diffusion metric and a benchmark alloy for comparison. As shown in \cref{fig:eigenspectrum}, the largest eigenvalue of $\mathbf{D}$, $\lambda_1$, corresponds to the vacancy tracer diffusion coefficient $D^*_{Va}$ and represents the fastest diffusion mode in the crystal. We use $D^*_{Va}$ as the primary scalar metric to quantify sluggish diffusion. The remaining eigenvalues scale to first order with vacancy concentration, so their sluggishness can be assessed by comparing the alloy vacancy concentration to the ROM prediction \cite{lee_modeling_2026}. When the vacancy concentrations in the real and hypothetical alloys are comparable, a similar analysis to that described below for $D^*_{Va}$ can be applied to the remaining eigenvalues. We compare $D^*_{Va}$ in the real alloy to that in a hypothetical single-component benchmark in which each site is perfectly disordered and the vacancy hop barrier is given by the ROM of the constituent elements. This comparison is quantified through an ``enhancement factor'' $\eta_{Va}$:
\begin{equation}
    \eta_{Va}=\frac{D^*_{Va,alloy}}{\langle D^*_{Va,pure} \rangle}.
\end{equation}
Here, $\langle D^*_{Va,pure} \rangle$ denotes the composition-weighted geometric mean of the vacancy tracer diffusion coefficients of the pure constituent elements. This quantity is conceptually similar to that introduced by \citeauthor{daw_sluggish_2021} \cite{daw_sluggish_2021}. A value of $\eta_{Va}>1$ indicates that the vacancy diffuses faster in the alloy than in the ROM benchmark, while $\eta_{Va}<1$ indicates sluggish diffusion. We additionally compute the vacancy correlation factor $f_{Va}$, which measures the efficiency of vacancy transport relative to an uncorrelated random walk. For a purely random walk $f_{Va}=1$, whereas $f_{Va}\to 0$ when successive hops tend to reverse the previous one.

\begin{figure*}
    \centering
    \includegraphics{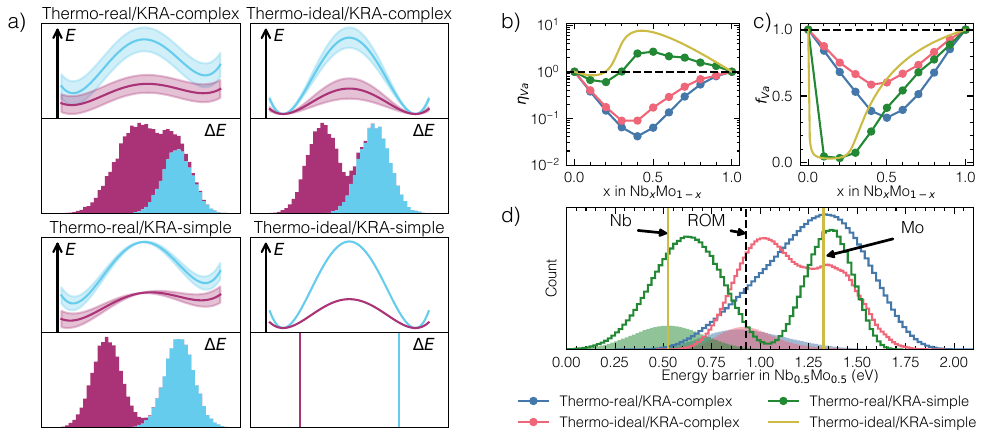}
    \caption{\textbf{a)} Schematic of four prototypical alloys used to disentangle thermodynamic and kinetic contributions to diffusion. Each panel shows the energy landscape for two hopping species, with shaded bands indicating environmental variability. Below each KRA energy landscape is the resulting distribution of KRA migration barriers. In the KRA-simple models, barriers are fixed at the pure-element values, while in the KRA-complex models they vary with local chemical environment. \textbf{b)} Enhancement factor $\eta_{Va}$ and \textbf{c)} vacancy correlation factor $f_{Va}$ as a function of Nb composition in Nb$_x$Mo$_{1-x}$ at 1500\,K for all four model alloys. The dashed line in \textbf{b)} marks $\eta_{Va} = 1$, separating sluggish ($\eta_{Va} < 1$) from anti-sluggish ($\eta_{Va} > 1$) behavior. \textbf{d)} Distribution of energy barriers encountered (unfilled) and accepted (filled) by the vacancy during kMC simulations at the equiatomic composition. Vertical solid lines mark the pure-element KRA values for Nb and Mo, and the dashed line indicates the rule-of-mixtures (ROM) estimate.}
    \label{fig:MoNb}
\end{figure*}

Beyond determining whether an alloy is sluggish, we seek to understand the origins of deviations from the ROM estimate. Thermodynamic factors affect the end-state energies of vacancy hops, so that regions where the vacancy has favorable interactions with surrounding elements can act as vacancy traps. Kinetic factors affect the migration barriers, where certain local environments or hopping species may have anomalously high or low barriers that lead to strongly correlated diffusion. To disentangle these contributions, we constructed four prototypical alloys that interpolate between ideal ROM behavior and the real alloy, as illustrated in \cref{fig:MoNb}a. A \textit{thermodynamically ideal} (thermo-ideal) model assigns identical end-state energies to all vacancy configurations, removing the possibility of vacancy traps. In the \textit{thermodynamically real} (thermo-real) model, end-state energies depend on the local chemical environment. Independently, a \textit{kinetically simple} (KRA-simple) model assigns constant migration barriers equal to the pure-element values for each hopping species, eliminating any distribution of saddle-point energies. In the \textit{kinetically complex} (KRA-complex) model, migration barriers vary with the local chemical environment. The transport coefficients in each of these four cases are compared against those of a thermodynamically and kinetically ideal alloy in which vacancy hop barriers are given by the ROM of the pure-element values.  For the thermo-ideal/KRA-simple case, diffusion metrics can be computed analytically using expressions derived in previous studies (\sicref{SI:sec:equations}) \cite{moleko_self-consistent_1989,manning_correlation_1971,mishin_monte_1997}.

We apply this framework to the binary Nb--Mo system at 1500~K, where Nb hops have on average a lower KRA barrier than Mo hops. \Cref{fig:MoNb}b shows $\eta_{Va}$ as a function of Nb composition for each of the four model variants. In the real alloy (blue curve), $\eta_{Va}<1$ across all compositions, indicating sluggish diffusion. The vacancy correlation factor, shown in \cref{fig:MoNb}c, is also below unity across the entire composition range, indicating significant correlation in vacancy transport relative to a random walk.

The thermo-ideal/KRA-complex alloy (red curves in \cref{fig:MoNb}b and c) exhibits $\eta_{Va}$ and $f_{Va}$ values very similar to those of the real alloy, with slightly less sluggish behavior and weaker vacancy correlation. This difference can be attributed to SRO. In the real alloy, SRO increases the number of Nb--Mo pairs relative to the ideal case, restricting vacancy motion because vacancies preferentially exchange with Nb atoms. The accepted and encountered energy barrier spectra at equiatomic composition are shown in \cref{fig:MoNb}d. Despite significant differences in the encountered barrier distributions between the two alloys, the accepted barrier spectra remain largely similar.

In contrast, the KRA-simple models produce distinctly different behavior from the KRA-complex alloy. For both the thermo-ideal and thermo-real cases, $\eta_{Va}$ and $f_{Va}$ deviate substantially from the kinetically complex cases (yellow and green curves in \cref{fig:MoNb}b and c). Anti-sluggish diffusion emerges when the Nb composition exceeds approximately 0.2 in the thermodynamically ideal alloy and around 0.3 in the thermodynamically real alloy, coinciding with a minimum in $f_{Va}$ (\cref{fig:MoNb}c). This trend is consistent with percolation behavior, as the site percolation threshold of the bcc lattice is 0.24 \cite{sykes_critical_1964}. Once a percolating network of fast Nb atoms is present in the disordered alloy, the vacancy can traverse the material through a connected series of low-barrier hops. In the thermodynamically real alloy, unfavorable SRO delays the onset of the percolating network of Nb atoms by reducing the number of nearest-neighbor Nb--Nb pairs. The distribution of accepted and encountered hops in the thermo-real/KRA-simple case (\cref{fig:MoNb}d) arises from variations in end-state energies, yet accepted hops remain centered around the Nb KRA barrier value, explaining the similar diffusion behavior between the thermo-ideal and thermo-real cases for the KRA-simple model.

Several conclusions can be drawn from the above analysis on the origins of sluggishness in these alloys. Thermodynamic effects appear to play a relatively minor role in governing diffusion, indicating that vacancy traps do not dominate transport in this class of materials \cite{tsai_sluggish_2013,zhang_ab_2022}. In contrast, kinetic barriers exert the dominant influence on vacancy transport, as evidenced by the substantially different diffusion behavior observed between the kinetically simple and kinetically complex cases. Anti-sluggish diffusion can also arise from percolation effects when the migration barrier spectrum is clearly separated between two species, allowing the species with lower barriers to form a percolating network that facilitates long-range vacancy transport. The results of \cref{fig:MoNb}c further suggest that the vacancy correlation factor alone does not provide a reliable indicator of sluggish diffusion. It is often assumed that sluggish diffusion in high-entropy alloys arises from a strong reduction of the vacancy correlation factor from unity. However, in \cref{fig:MoNb}, similar values of $f_{Va}$ can correspond to substantially different $\eta_{Va}$ values, indicating that correlation effects alone cannot fully capture the underlying diffusion behavior.

\begin{figure*}
    \centering
    \includegraphics{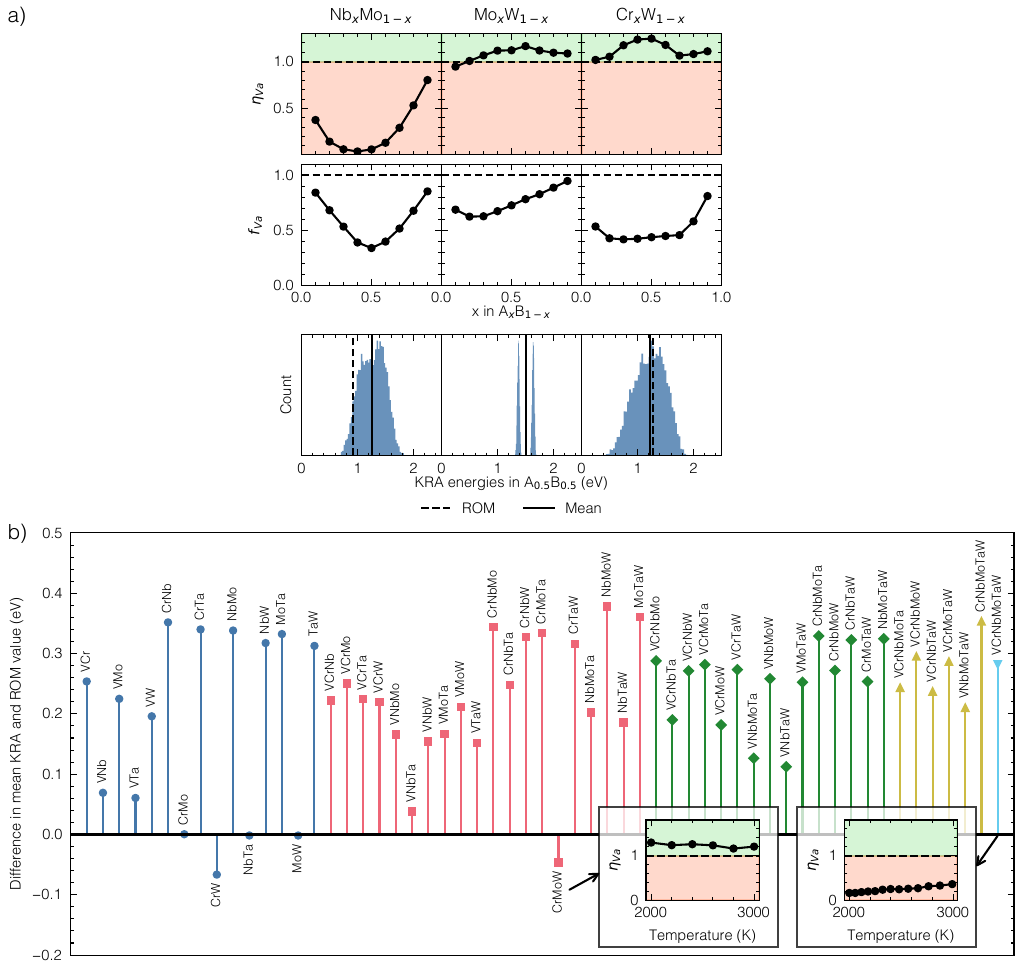}
    \caption{\textbf{a)} Enhancement factor $\eta_{Va}$ (top), vacancy correlation factor $f_{Va}$ (middle), and KRA energy spectrum at equiatomic composition (bottom) for three binary alloys: Nb$_x$Mo$_{1-x}$, Mo$_x$W$_{1-x}$, and Cr$_x$W$_{1-x}$. Green and red shading in the $\eta_{Va}$ panels indicate anti-sluggish ($\eta_{Va} > 1$) and sluggish ($\eta_{Va} < 1$) regimes, respectively. In the KRA spectra, the dashed line marks the ROM estimate and the solid line marks the mean of the distribution. \textbf{b)} Difference between the mean KRA of the vacancy barrier spectrum and the ROM prediction for all 57 equiatomic alloys in the V--Cr--Nb--Mo--Ta--W composition space, grouped by chemical complexity (binaries through the senary). Alloys with negative values are candidates for anti-sluggish diffusion. Insets show the temperature-dependent enhancement factor $\eta_{Va}$ for CrMoW (anti-sluggish) and VCrNbMoTaW (sluggish), verified by kMC simulations.}
    \label{fig:alloys_and_screening}
\end{figure*}

Having established the distribution of KRA barriers as a primary factor controlling vacancy diffusion in the Nb--Mo alloy, we computed the enhancement factor $\eta_{Va}$ and the vacancy correlation factor $f_{Va}$ for two additional binary alloys, Mo--W and Cr--W, at 1500~K and 2500~K, respectively, where the alloys are stable in the disordered phase. \Cref{fig:alloys_and_screening}a shows the vacancy diffusion metrics for Nb--Mo, Cr--W and Mo--W, with KRA and end-state energies computed from the real interactions learned from our first-principles electronic structure calculations. NbMo exhibits sluggish diffusion ($\eta_{Va}<1$) across the entire compositional range, with enhancement factors as low as 0.1. In contrast, MoW displays anti-sluggish behavior, with $\eta_{Va}$ falling below unity only when the Mo composition is below roughly 0.2, and CrW exhibits anti-sluggish diffusion ($\eta_{Va}>1$) throughout the composition range.

\Cref{fig:alloys_and_screening}a also plots the KRA energy spectrum at the equiatomic composition for all three alloys in the random alloy limit. In NbMo, the KRA distribution is approximately normal with a mean above the ROM value, suggesting that vacancies encounter larger barriers on average and leading to sluggish diffusion. In MoW, the KRA distribution separates into two distinct and relatively narrow peaks corresponding to Mo and W hops, with Mo having lower barriers. This resembles the thermo-ideal/KRA-simple case shown in \cref{fig:MoNb}. The mean of this distribution closely matches the ROM prediction, and the observed anti-sluggish behavior can be rationalized via percolation. Once a percolating network of Mo atoms forms in the disordered state, the vacancy rapidly diffuses through it. For CrW, the mean of the vacancy KRA spectrum lies below the ROM estimate, indicating lower average barriers and explaining the observed anti-sluggish diffusion ($\eta_{Va} > 1$).

Taken together, the results of \cref{fig:MoNb,fig:alloys_and_screening}a indicate two distinct origins of sluggish and anti-sluggish diffusion. The first applies when the barrier spectrum resolves into separable fast and slow atoms, as in the KRA-simple cases of \cref{fig:MoNb} and in Mo--W (\cref{fig:alloys_and_screening}a). Here a percolation description applies, in which sites occupied by the fast species form a connected network once their fraction exceeds the site percolation threshold (0.24 for bcc \cite{sykes_critical_1964}), and the vacancy traverses the material through fast hops. The presence or absence of sluggish diffusion can then be gauged by whether a percolating network of the fast diffuser exists.
When a clearly resolved multimodal spectrum is absent, this distinction cannot be drawn and the percolation description does not apply. In this case, the mean of the vacancy KRA spectrum, relative to the ROM estimate, sets the diffusion behavior. When the mean exceeds the ROM estimate, as in Nb--Mo, the system is sluggish. When it falls below, as in Cr--W, anti-sluggish behavior emerges across the full composition range.

The observations of \cref{fig:MoNb,fig:alloys_and_screening}a suggest a simple heuristic for predicting alloy sluggishness, namely evaluating the shape of the vacancy barrier spectrum and the relative position of the ROM value within it. Our eLCE formalism enables rapid evaluation of KRA values for arbitrary configurations, allowing efficient screening across large composition spaces. We applied this heuristic to all 57 possible equiatomic compositions in the senary space, comparing the mean of the KRA spectrum with the rule-of-mixtures prediction. The difference is shown in \cref{fig:alloys_and_screening}b. Only four alloys exhibit a mean KRA below the ROM value, suggesting potential anti-sluggish behavior. Of these, NbTa and MoW show near-ideal behavior, while CrW and CrMoW deviate more substantially. Visual inspection further shows that, aside from NbTa and MoW, no alloys display a spectrum resembling the thermo-ideal/KRA-simple case with a multi-modal distribution of KRAs.

The predictions of sluggishness or anti-sluggishness based on the proposed heuristic are verified through rigorous kMC simulations with our eLCE model. As shown in the insets of \cref{fig:alloys_and_screening}b, finite-temperature simulations across a range of temperatures for the CrMoW alloy predict anti-sluggish diffusion with $\eta_{Va}>1$. In contrast, in the equiatomic senary VCrNbMoTaW alloy, where the typical vacancy KRA is nearly 0.3\,eV above the ROM value, the enhancement factor is strongly suppressed, approaching values of $\approx 0.1$ at low temperatures. The results of \cref{fig:alloys_and_screening}b validate the heuristic for screening alloys for sluggish diffusion, demonstrate that our results extend beyond binary systems, and indicate that most alloys in this senary composition space remain sluggish, in contrast to the findings reported by \citeauthor{daw_sluggish_2021} for fcc alloys \cite{daw_sluggish_2021}.

Phase-separating systems (\textit{e.g.}, VNb, VTa, CrMo) tend to have low or negative differences compared to miscible systems. These findings suggest a correlation between bonding energetics and diffusion barriers, in which stronger bonding increases the energy barriers for vacancy hops. Other descriptors, such as lattice mismatch, have also been proposed as a potential correlating factor \cite{daw_sluggish_2021}. However, no clear correlation is observed in our results (\sicref{SI:fig:lattice_mismatch}). Phase-separating systems also tend to have higher vacancy concentrations than the ROM estimate \cite{lee_modeling_2026}, and the smaller eigenvalues of $\mathbf{D}$ in these alloys are likely to be enhanced as well due to elevated vacancy concentrations.

While this screening approach provides a computationally inexpensive indicator of diffusion trends, a rigorous assessment requires a full kinetic analysis. In particular, eLCE kinetic models alongside eCE energy models combined with kMC simulations are necessary to capture collective effects such as correlation factors, percolation pathways, short-range order, and the full spectrum of local environments encountered during vacancy migration. Nevertheless, the screening already highlights that sluggish diffusion is not exclusive to a ``high-entropy effect'' \cite{tsai_sluggish_2013,hsu_clarifying_2024}, as several binaries and ternaries display strong sluggish characteristics.

The methodology described in this study neglects vibrational contributions to both the vacancy formation free energy and the KRA values, and assumes the dilute-vacancy limit, ignoring the formation of divacancies or larger vacancy clusters. Vibrational contributions have been shown to be especially important at temperatures approaching the melting point of the alloy \cite{zhang_ab_2025} or in the presence of a large concentration of group 4 elements \cite{sangiovanni_superioniclike_2019}. We also neglect thermal expansion and restrict vacancy migration to nearest-neighbor hops. More complex mechanisms, such as next-nearest-neighbor hops or ring-like exchange processes, may contribute to diffusion in multicomponent alloys but are not captured in the current model. Despite these assumptions inherent to on-lattice models, the present framework is expected to robustly reproduce chemical and diffusion trends. We benchmarked our results against the limited experimental data available for concentrated alloys within the six-component space, namely the Mo--W alloy. \Cref{SI:fig:experimental} in the SI compares the Mo enhancement factor $\eta_{Mo}$ in the Mo--W alloy across the full composition range, as measured experimentally \cite{larikov_notitle_1967,neumann_self-diffusion_2008,erley_diffusion_1974} and as predicted by kMC simulations using the eLCE model parameterized in this work. Predicted and measured enhancement factors agree to within a few percent for $x_{Mo} \gtrsim 0.4$, with the largest deviation, approximately 25\%, near $x_{Mo} \approx 0.2$, and both place $\eta_{Mo}$ above unity across the composition range. This agreement is achieved with no fitting parameters, indicating that the model quantitatively captures non-dilute diffusion in this system. Discrepancies in the absolute magnitude of the tracer diffusion coefficients between computation and experiment can be largely accounted for by including vibrational contributions to the vacancy formation free energy together with an empirical term that absorbs residual errors from the DFT energies and the cluster expansion fit (\cref{SI:sec:benchmarks} in the SI). We additionally benchmarked the impact of the approximations above and found that vibrational contributions primarily introduce systematic shifts in the absolute diffusion coefficients. As our analysis of sluggish diffusion is based on enhancement factors relative to the rule-of-mixtures estimate, our benchmark confirms that these shifts largely cancel and have only a minor impact on the reported enhancement factors. Thermal expansion modifies the enhancement factor for Mo--W by approximately 0.5\%, negligible relative to the larger variations in $\eta_{Va}$ discussed in this work. More broadly, these approximations reflect the current state of the art in first-principles diffusion modeling of disordered alloys. We are not aware of a computationally tractable framework that simultaneously incorporates chemical disorder, lattice relaxation, anharmonic vibrations, magnetic disorder, and long-timescale diffusion kinetics within a single predictive model.

While our analysis establishes the dominant role of kinetic barriers in groups 5–6 refractory alloys, applying the same decomposition to other alloy classes, including systems with stronger thermodynamic vacancy trapping or persistent magnetic ordering, is a worthwhile direction for future work. Finally, future studies should analyze the full eigenspectrum of the diffusion matrix to explore the relationship between experimentally measurable diffusion metrics and the individual elements of the diffusion matrix in complex concentrated alloys. 

In conclusion, we have developed and applied the eLCE framework to compute vacancy-mediated diffusion in multicomponent alloys from first principles, enabling quantitatively accurate evaluation of transport in compositionally complex systems at experimentally relevant timescales. Our analysis reveals that diffusion behavior is primarily governed by the distribution of local migration barriers rather than thermodynamic effects or the vacancy correlation factor alone. This insight leads to a simple heuristic based on the position of the rule-of-mixtures prediction within the vacancy barrier spectrum, which we use to rapidly screen all equiatomic compositions in the senary V--Cr--Nb--Mo--Ta--W space. The screening demonstrates that sluggish diffusion is not exclusive to high-entropy compositions, as several binaries and ternaries already display strong sluggish characteristics. More broadly, the eLCE framework establishes a predictive, first-principles approach for exploring transport phenomena in a wide range of complex materials utilized in structural, functional and energy storage applications.

\section{Methods}
\label{sec:methods}
\subsection{Diffusion barrier calculation}
We computed diffusion barriers across the senary composition space using a two-step workflow that achieves near-DFT accuracy at reduced computational cost. In the first step, end-state configurations were relaxed using the MACE-MPA0 interatomic potential \cite{batatia_mace_2023}. Three intermediate images were generated by linear interpolation between the relaxed end states, and climbing-image nudged elastic band (CI-NEB) calculations \cite{henkelman_climbing_2000, henkelman_improved_2000} were performed with the same MACE potential. In the second step, the energies of both end states and the corresponding transition states were recomputed via single-point DFT calculations. All calculations used a $4 \times 4 \times 4$ supercell of the conventional bcc unit cell (128 atoms). Exchange-correlation effects were treated with the Perdew-Burke-Ernzerhof (PBE) functional \cite{perdew_generalized_1996} and core electrons were described using projector augmented wave (PAW) potentials \cite{kresse_efficiency_1996}. The plane-wave energy cutoff was set to 550\,eV, and Brillouin zone sampling employed a $\Gamma$-centered Monkhorst-Pack $k$-point mesh with a grid density of 55\,\AA{} \cite{monkhorst_special_1976}. A spring constant of 0.1\,eV/\AA{} was used in all CI-NEB calculations, with forces converged to within 0.01\,eV/\AA{}.

We validated this workflow against full DFT-based CI-NEB calculations for a subset of 31 configurations spanning the binary to senary space. The MAE between KRA values obtained from the MACE-based workflow and the full DFT CI-NEB calculations is 36.6\,meV (\sicref{SI:fig:parity_plot_mace_dft}), well below both the eLCE model error and the intrinsic uncertainty of transition-state DFT calculations \cite{van_der_ven_rechargeable_2020}.

The training dataset was generated through an iterative active-learning workflow. In the first iteration, equiatomic compositions spanning the binary to senary space were sampled, with vacancy hops selected uniformly across hopping species. An initial eLCE model was trained on this dataset. In subsequent iterations, random compositions containing a single vacancy were generated across the same compositional space, and canonical Monte Carlo simulations at 500, 1000, and 2000\,K were used to produce thermodynamically relevant configurations. Vacancy hop barriers were estimated using the current eLCE model, and configurations were sampled uniformly across the full barrier spectrum to ensure that both low- and high-barrier events were well represented. The model was retrained after each iteration until the error converged.

\subsection{Embedded local cluster expansions}
The eLCE was implemented in \texttt{Python} using the \texttt{PyTorch} library. Point and pair clusters surrounding the hop cluster were constructed with maximum cutoff distances of 8 and 5\,\AA{}, respectively, measured from the hop cluster. Each cluster was augmented by including the hop cluster itself. The final model comprised 1 empty cluster, 21 point clusters, 26 pair clusters, and 7 triplet clusters. Occupational and Chebyshev site basis functions were used for the hop cluster and local environment, respectively. These basis functions were projected into a lower-dimensional space through a learnable embedding matrix, and symmetrized cluster functions constructed as tensor products of the embedded basis functions served as input to a 4-layer neural network (32 $\times$ 32 $\times$ 8 $\times$ 1) with ReLU activations in the hidden layers and a linear output layer. The training loss was minimized using the Adam optimizer with a learning rate scheduler. Overfitting was controlled through early stopping on the validation set and L1/L2 regularization. The embedding matrix was re-normalized after each iteration and initialized according to the chemical properties of each element following the approach of \cite{muller_constructing_2025}. A separate 3-eCE energy model was trained on the V--Cr--Nb--Mo--Ta--W--Va composition space using data from \citeauthor{lee_modeling_2026}\cite{lee_modeling_2026}, achieving a test MAE of 3.5\,meV/site.

\subsection{Kinetic Monte Carlo simulations}
Rejection-free kMC simulations \cite{van_der_ven_first-principles_2001} were performed on a $10\times10\times10$ supercell of the conventional bcc structure (2000 atoms) with the vibrational prefactor $\nu^*$ in \cref{eqn:transition_state} set to 1\,THz. For each calculation, 500 distinct initial configurations were generated using canonical Monte Carlo, followed by 50,000 and 100,000 kMC steps for the binary and senary alloys, respectively, yielding 25 and 50 million total kMC steps per calculation.

\section{Data availability}
The data used in the present work are available through the MaterialsCloud (A link to the repository will be made available after the acceptance of the paper). Source Data are provided with this paper.

\section{Acknowledgments}
We acknowledge access to Eiger at the Swiss National Supercomputing Centre with project ID mr30, provided with support from NCCR MARVEL, a National Centre of Competence in Research, funded by the Swiss National Science Foundation, Switzerland (grant number 205602). We also acknowledge support from the SNSF through project number 215178.

\bibliography{references}

\begin{thebibliography}{75}%
\makeatletter
\providecommand \@ifxundefined [1]{%
 \@ifx{#1\undefined}
}%
\providecommand \@ifnum [1]{%
 \ifnum #1\expandafter \@firstoftwo
 \else \expandafter \@secondoftwo
 \fi
}%
\providecommand \@ifx [1]{%
 \ifx #1\expandafter \@firstoftwo
 \else \expandafter \@secondoftwo
 \fi
}%
\providecommand \natexlab [1]{#1}%
\providecommand \enquote  [1]{``#1''}%
\providecommand \bibnamefont  [1]{#1}%
\providecommand \bibfnamefont [1]{#1}%
\providecommand \citenamefont [1]{#1}%
\providecommand \href@noop [0]{\@secondoftwo}%
\providecommand \href [0]{\begingroup \@sanitize@url \@href}%
\providecommand \@href[1]{\@@startlink{#1}\@@href}%
\providecommand \@@href[1]{\endgroup#1\@@endlink}%
\providecommand \@sanitize@url [0]{\catcode `\\12\catcode `\$12\catcode `\&12\catcode `\#12\catcode `\^12\catcode `\_12\catcode `\%12\relax}%
\providecommand \@@startlink[1]{}%
\providecommand \@@endlink[0]{}%
\providecommand \url  [0]{\begingroup\@sanitize@url \@url }%
\providecommand \@url [1]{\endgroup\@href {#1}{\urlprefix }}%
\providecommand \urlprefix  [0]{URL }%
\providecommand \Eprint [0]{\href }%
\providecommand \doibase [0]{https://doi.org/}%
\providecommand \selectlanguage [0]{\@gobble}%
\providecommand \bibinfo  [0]{\@secondoftwo}%
\providecommand \bibfield  [0]{\@secondoftwo}%
\providecommand \translation [1]{[#1]}%
\providecommand \BibitemOpen [0]{}%
\providecommand \bibitemStop [0]{}%
\providecommand \bibitemNoStop [0]{.\EOS\space}%
\providecommand \EOS [0]{\spacefactor3000\relax}%
\providecommand \BibitemShut  [1]{\csname bibitem#1\endcsname}%
\let\auto@bib@innerbib\@empty
\bibitem [{\citenamefont {Ardell}\ and\ \citenamefont {Ozolins}(2005)}]{ardell_trans-interface_2005}%
  \BibitemOpen
  \bibfield  {author} {\bibinfo {author} {\bibfnamefont {A.~J.}\ \bibnamefont {Ardell}}\ and\ \bibinfo {author} {\bibfnamefont {V.}~\bibnamefont {Ozolins}},\ }\bibfield  {title} {\bibinfo {title} {Trans-interface diffusion-controlled coarsening},\ }\href {https://doi.org/10.1038/nmat1340} {\bibfield  {journal} {\bibinfo  {journal} {Nature Materials}\ }\textbf {\bibinfo {volume} {4}},\ \bibinfo {pages} {309} (\bibinfo {year} {2005})}\BibitemShut {NoStop}%
\bibitem [{\citenamefont {Miracle}\ and\ \citenamefont {Senkov}(2017)}]{miracle_critical_2017}%
  \BibitemOpen
  \bibfield  {author} {\bibinfo {author} {\bibfnamefont {D.~B.}\ \bibnamefont {Miracle}}\ and\ \bibinfo {author} {\bibfnamefont {O.~N.}\ \bibnamefont {Senkov}},\ }\bibfield  {title} {\bibinfo {title} {A critical review of high entropy alloys and related concepts},\ }\href {https://doi.org/10.1016/j.actamat.2016.08.081} {\bibfield  {journal} {\bibinfo  {journal} {Acta Materialia}\ }\textbf {\bibinfo {volume} {122}},\ \bibinfo {pages} {448} (\bibinfo {year} {2017})}\BibitemShut {NoStop}%
\bibitem [{\citenamefont {Darling}\ \emph {et~al.}(2016)\citenamefont {Darling}, \citenamefont {Rajagopalan}, \citenamefont {Komarasamy}, \citenamefont {Bhatia}, \citenamefont {Hornbuckle}, \citenamefont {Mishra},\ and\ \citenamefont {Solanki}}]{darling_extreme_2016}%
  \BibitemOpen
  \bibfield  {author} {\bibinfo {author} {\bibfnamefont {K.~A.}\ \bibnamefont {Darling}}, \bibinfo {author} {\bibfnamefont {M.}~\bibnamefont {Rajagopalan}}, \bibinfo {author} {\bibfnamefont {M.}~\bibnamefont {Komarasamy}}, \bibinfo {author} {\bibfnamefont {M.~A.}\ \bibnamefont {Bhatia}}, \bibinfo {author} {\bibfnamefont {B.~C.}\ \bibnamefont {Hornbuckle}}, \bibinfo {author} {\bibfnamefont {R.~S.}\ \bibnamefont {Mishra}},\ and\ \bibinfo {author} {\bibfnamefont {K.~N.}\ \bibnamefont {Solanki}},\ }\bibfield  {title} {\bibinfo {title} {Extreme creep resistance in a microstructurally stable nanocrystalline alloy},\ }\href {https://doi.org/10.1038/nature19313} {\bibfield  {journal} {\bibinfo  {journal} {Nature}\ }\textbf {\bibinfo {volume} {537}},\ \bibinfo {pages} {378} (\bibinfo {year} {2016})}\BibitemShut {NoStop}%
\bibitem [{\citenamefont {Hinrichs}\ \emph {et~al.}(2025)\citenamefont {Hinrichs}, \citenamefont {Winkens}, \citenamefont {Kramer}, \citenamefont {Falcão}, \citenamefont {Hahn}, \citenamefont {Schliephake}, \citenamefont {Eusterholz}, \citenamefont {Sen}, \citenamefont {Galetz}, \citenamefont {Inui}, \citenamefont {Kauffmann},\ and\ \citenamefont {Heilmaier}}]{hinrichs_ductile_2025}%
  \BibitemOpen
  \bibfield  {author} {\bibinfo {author} {\bibfnamefont {F.}~\bibnamefont {Hinrichs}}, \bibinfo {author} {\bibfnamefont {G.}~\bibnamefont {Winkens}}, \bibinfo {author} {\bibfnamefont {L.~K.}\ \bibnamefont {Kramer}}, \bibinfo {author} {\bibfnamefont {G.}~\bibnamefont {Falcão}}, \bibinfo {author} {\bibfnamefont {E.~M.}\ \bibnamefont {Hahn}}, \bibinfo {author} {\bibfnamefont {D.}~\bibnamefont {Schliephake}}, \bibinfo {author} {\bibfnamefont {M.~K.}\ \bibnamefont {Eusterholz}}, \bibinfo {author} {\bibfnamefont {S.}~\bibnamefont {Sen}}, \bibinfo {author} {\bibfnamefont {M.~C.}\ \bibnamefont {Galetz}}, \bibinfo {author} {\bibfnamefont {H.}~\bibnamefont {Inui}}, \bibinfo {author} {\bibfnamefont {A.}~\bibnamefont {Kauffmann}},\ and\ \bibinfo {author} {\bibfnamefont {M.}~\bibnamefont {Heilmaier}},\ }\bibfield  {title} {\bibinfo {title} {A ductile chromium–molybdenum alloy resistant to high-temperature oxidation},\ }\href {https://doi.org/10.1038/s41586-025-09516-8} {\bibfield  {journal} {\bibinfo  {journal}
  {Nature}\ }\textbf {\bibinfo {volume} {646}},\ \bibinfo {pages} {331} (\bibinfo {year} {2025})}\BibitemShut {NoStop}%
\bibitem [{\citenamefont {Wang}\ \emph {et~al.}(2025)\citenamefont {Wang}, \citenamefont {Kwon}, \citenamefont {Oh}, \citenamefont {Lee}, \citenamefont {Yun}, \citenamefont {Lee}, \citenamefont {Seo}, \citenamefont {Yoo}, \citenamefont {Jeong}, \citenamefont {Kim},\ and\ \citenamefont {Lee}}]{wang_multiscale_2025}%
  \BibitemOpen
  \bibfield  {author} {\bibinfo {author} {\bibfnamefont {J.}~\bibnamefont {Wang}}, \bibinfo {author} {\bibfnamefont {H.}~\bibnamefont {Kwon}}, \bibinfo {author} {\bibfnamefont {S.-H.}\ \bibnamefont {Oh}}, \bibinfo {author} {\bibfnamefont {J.~H.}\ \bibnamefont {Lee}}, \bibinfo {author} {\bibfnamefont {D.~W.}\ \bibnamefont {Yun}}, \bibinfo {author} {\bibfnamefont {H.}~\bibnamefont {Lee}}, \bibinfo {author} {\bibfnamefont {S.-M.}\ \bibnamefont {Seo}}, \bibinfo {author} {\bibfnamefont {Y.-S.}\ \bibnamefont {Yoo}}, \bibinfo {author} {\bibfnamefont {H.~W.}\ \bibnamefont {Jeong}}, \bibinfo {author} {\bibfnamefont {H.~S.}\ \bibnamefont {Kim}},\ and\ \bibinfo {author} {\bibfnamefont {B.-J.}\ \bibnamefont {Lee}},\ }\bibfield  {title} {\bibinfo {title} {Multiscale computational framework linking alloy composition to microstructure evolution via machine learning and nanoscale analysis},\ }\href {https://doi.org/10.1038/s41524-025-01730-2} {\bibfield  {journal} {\bibinfo  {journal} {npj Computational Materials}\ }\textbf
  {\bibinfo {volume} {11}},\ \bibinfo {pages} {230} (\bibinfo {year} {2025})}\BibitemShut {NoStop}%
\bibitem [{\citenamefont {Chen}\ \emph {et~al.}(2024)\citenamefont {Chen}, \citenamefont {Lun}, \citenamefont {Zhao}, \citenamefont {Koirala}, \citenamefont {Li}, \citenamefont {Sun}, \citenamefont {O’Keefe}, \citenamefont {Yang}, \citenamefont {Cai}, \citenamefont {Wang}, \citenamefont {Ji}, \citenamefont {Grey}, \citenamefont {Ouyang},\ and\ \citenamefont {Ceder}}]{chen_unlocking_2024}%
  \BibitemOpen
  \bibfield  {author} {\bibinfo {author} {\bibfnamefont {Y.}~\bibnamefont {Chen}}, \bibinfo {author} {\bibfnamefont {Z.}~\bibnamefont {Lun}}, \bibinfo {author} {\bibfnamefont {X.}~\bibnamefont {Zhao}}, \bibinfo {author} {\bibfnamefont {K.~P.}\ \bibnamefont {Koirala}}, \bibinfo {author} {\bibfnamefont {L.}~\bibnamefont {Li}}, \bibinfo {author} {\bibfnamefont {Y.}~\bibnamefont {Sun}}, \bibinfo {author} {\bibfnamefont {C.~A.}\ \bibnamefont {O’Keefe}}, \bibinfo {author} {\bibfnamefont {X.}~\bibnamefont {Yang}}, \bibinfo {author} {\bibfnamefont {Z.}~\bibnamefont {Cai}}, \bibinfo {author} {\bibfnamefont {C.}~\bibnamefont {Wang}}, \bibinfo {author} {\bibfnamefont {H.}~\bibnamefont {Ji}}, \bibinfo {author} {\bibfnamefont {C.~P.}\ \bibnamefont {Grey}}, \bibinfo {author} {\bibfnamefont {B.}~\bibnamefont {Ouyang}},\ and\ \bibinfo {author} {\bibfnamefont {G.}~\bibnamefont {Ceder}},\ }\bibfield  {title} {\bibinfo {title} {Unlocking {Li} superionic conductivity in face-centred cubic oxides via face-sharing
  configurations},\ }\href {https://doi.org/10.1038/s41563-024-01800-8} {\bibfield  {journal} {\bibinfo  {journal} {Nature Materials}\ }\textbf {\bibinfo {volume} {23}},\ \bibinfo {pages} {535} (\bibinfo {year} {2024})}\BibitemShut {NoStop}%
\bibitem [{\citenamefont {Jun}\ \emph {et~al.}(2024)\citenamefont {Jun}, \citenamefont {Chen}, \citenamefont {Wei}, \citenamefont {Yang},\ and\ \citenamefont {Ceder}}]{jun_diffusion_2024}%
  \BibitemOpen
  \bibfield  {author} {\bibinfo {author} {\bibfnamefont {K.}~\bibnamefont {Jun}}, \bibinfo {author} {\bibfnamefont {Y.}~\bibnamefont {Chen}}, \bibinfo {author} {\bibfnamefont {G.}~\bibnamefont {Wei}}, \bibinfo {author} {\bibfnamefont {X.}~\bibnamefont {Yang}},\ and\ \bibinfo {author} {\bibfnamefont {G.}~\bibnamefont {Ceder}},\ }\bibfield  {title} {\bibinfo {title} {Diffusion mechanisms of fast lithium-ion conductors},\ }\href {https://doi.org/10.1038/s41578-024-00715-9} {\bibfield  {journal} {\bibinfo  {journal} {Nature Reviews Materials}\ }\textbf {\bibinfo {volume} {9}},\ \bibinfo {pages} {887} (\bibinfo {year} {2024})}\BibitemShut {NoStop}%
\bibitem [{\citenamefont {Chmelik}\ and\ \citenamefont {Kärger}(2010)}]{chmelik_situ_2010}%
  \BibitemOpen
  \bibfield  {author} {\bibinfo {author} {\bibfnamefont {C.}~\bibnamefont {Chmelik}}\ and\ \bibinfo {author} {\bibfnamefont {J.}~\bibnamefont {Kärger}},\ }\bibfield  {title} {\bibinfo {title} {In situ study on molecular diffusion phenomena in nanoporous catalytic solids},\ }\href {https://doi.org/10.1039/C0CS00100G} {\bibfield  {journal} {\bibinfo  {journal} {Chemical Society Reviews}\ }\textbf {\bibinfo {volume} {39}},\ \bibinfo {pages} {4864} (\bibinfo {year} {2010})}\BibitemShut {NoStop}%
\bibitem [{\citenamefont {Panda}\ \emph {et~al.}(2025)\citenamefont {Panda}, \citenamefont {Maity}, \citenamefont {Sarkar}, \citenamefont {Manna}, \citenamefont {Mondal},\ and\ \citenamefont {Haldar}}]{panda_diffusion-programmed_2025}%
  \BibitemOpen
  \bibfield  {author} {\bibinfo {author} {\bibfnamefont {S.}~\bibnamefont {Panda}}, \bibinfo {author} {\bibfnamefont {T.}~\bibnamefont {Maity}}, \bibinfo {author} {\bibfnamefont {S.}~\bibnamefont {Sarkar}}, \bibinfo {author} {\bibfnamefont {A.~K.}\ \bibnamefont {Manna}}, \bibinfo {author} {\bibfnamefont {J.}~\bibnamefont {Mondal}},\ and\ \bibinfo {author} {\bibfnamefont {R.}~\bibnamefont {Haldar}},\ }\bibfield  {title} {\bibinfo {title} {Diffusion-programmed catalysis in nanoporous material},\ }\href {https://doi.org/10.1038/s41467-025-56575-6} {\bibfield  {journal} {\bibinfo  {journal} {Nature Communications}\ }\textbf {\bibinfo {volume} {16}},\ \bibinfo {pages} {1231} (\bibinfo {year} {2025})}\BibitemShut {NoStop}%
\bibitem [{\citenamefont {Dahan}\ \emph {et~al.}(2021)\citenamefont {Dahan}, \citenamefont {Baranovskiy}, \citenamefont {Natanzon},\ and\ \citenamefont {Amouyal}}]{dahan_first-principles_2021}%
  \BibitemOpen
  \bibfield  {author} {\bibinfo {author} {\bibfnamefont {M.~H.}\ \bibnamefont {Dahan}}, \bibinfo {author} {\bibfnamefont {A.}~\bibnamefont {Baranovskiy}}, \bibinfo {author} {\bibfnamefont {Y.}~\bibnamefont {Natanzon}},\ and\ \bibinfo {author} {\bibfnamefont {Y.}~\bibnamefont {Amouyal}},\ }\bibfield  {title} {\bibinfo {title} {A first-principles study of the temperature-dependent diffusion coefficients of silver in the thermoelectric compound {PbTe}},\ }\href {https://doi.org/10.1016/j.actamat.2020.10.060} {\bibfield  {journal} {\bibinfo  {journal} {Acta Materialia}\ }\textbf {\bibinfo {volume} {202}},\ \bibinfo {pages} {243} (\bibinfo {year} {2021})}\BibitemShut {NoStop}%
\bibitem [{\citenamefont {Zhang}\ \emph {et~al.}(2021)\citenamefont {Zhang}, \citenamefont {Roth}, \citenamefont {Tolborg}, \citenamefont {Takahashi}, \citenamefont {Song}, \citenamefont {Bondesgaard}, \citenamefont {Nishibori},\ and\ \citenamefont {Iversen}}]{zhang_direct_2021}%
  \BibitemOpen
  \bibfield  {author} {\bibinfo {author} {\bibfnamefont {J.}~\bibnamefont {Zhang}}, \bibinfo {author} {\bibfnamefont {N.}~\bibnamefont {Roth}}, \bibinfo {author} {\bibfnamefont {K.}~\bibnamefont {Tolborg}}, \bibinfo {author} {\bibfnamefont {S.}~\bibnamefont {Takahashi}}, \bibinfo {author} {\bibfnamefont {L.}~\bibnamefont {Song}}, \bibinfo {author} {\bibfnamefont {M.}~\bibnamefont {Bondesgaard}}, \bibinfo {author} {\bibfnamefont {E.}~\bibnamefont {Nishibori}},\ and\ \bibinfo {author} {\bibfnamefont {B.~B.}\ \bibnamefont {Iversen}},\ }\bibfield  {title} {\bibinfo {title} {Direct observation of one-dimensional disordered diffusion channel in a chain-like thermoelectric with ultralow thermal conductivity},\ }\href {https://doi.org/10.1038/s41467-021-27007-y} {\bibfield  {journal} {\bibinfo  {journal} {Nature Communications}\ }\textbf {\bibinfo {volume} {12}},\ \bibinfo {pages} {6709} (\bibinfo {year} {2021})}\BibitemShut {NoStop}%
\bibitem [{\citenamefont {Yeh}\ \emph {et~al.}(2004)\citenamefont {Yeh}, \citenamefont {Chen}, \citenamefont {Lin}, \citenamefont {Gan}, \citenamefont {Chin}, \citenamefont {Shun}, \citenamefont {Tsau},\ and\ \citenamefont {Chang}}]{yeh_nanostructured_2004}%
  \BibitemOpen
  \bibfield  {author} {\bibinfo {author} {\bibfnamefont {J.-W.}\ \bibnamefont {Yeh}}, \bibinfo {author} {\bibfnamefont {S.-K.}\ \bibnamefont {Chen}}, \bibinfo {author} {\bibfnamefont {S.-J.}\ \bibnamefont {Lin}}, \bibinfo {author} {\bibfnamefont {J.-Y.}\ \bibnamefont {Gan}}, \bibinfo {author} {\bibfnamefont {T.-S.}\ \bibnamefont {Chin}}, \bibinfo {author} {\bibfnamefont {T.-T.}\ \bibnamefont {Shun}}, \bibinfo {author} {\bibfnamefont {C.-H.}\ \bibnamefont {Tsau}},\ and\ \bibinfo {author} {\bibfnamefont {S.-Y.}\ \bibnamefont {Chang}},\ }\bibfield  {title} {\bibinfo {title} {Nanostructured {High}-{Entropy} {Alloys} with {Multiple} {Principal} {Elements}: {Novel} {Alloy} {Design} {Concepts} and {Outcomes}},\ }\href {https://doi.org/10.1002/adem.200300567} {\bibfield  {journal} {\bibinfo  {journal} {Advanced Engineering Materials}\ }\textbf {\bibinfo {volume} {6}},\ \bibinfo {pages} {299} (\bibinfo {year} {2004})}\BibitemShut {NoStop}%
\bibitem [{\citenamefont {Tsai}\ \emph {et~al.}(2013)\citenamefont {Tsai}, \citenamefont {Tsai},\ and\ \citenamefont {Yeh}}]{tsai_sluggish_2013}%
  \BibitemOpen
  \bibfield  {author} {\bibinfo {author} {\bibfnamefont {K.~Y.}\ \bibnamefont {Tsai}}, \bibinfo {author} {\bibfnamefont {M.~H.}\ \bibnamefont {Tsai}},\ and\ \bibinfo {author} {\bibfnamefont {J.~W.}\ \bibnamefont {Yeh}},\ }\bibfield  {title} {\bibinfo {title} {Sluggish diffusion in {Co}–{Cr}–{Fe}–{Mn}–{Ni} high-entropy alloys},\ }\href {https://doi.org/10.1016/j.actamat.2013.04.058} {\bibfield  {journal} {\bibinfo  {journal} {Acta Materialia}\ }\textbf {\bibinfo {volume} {61}},\ \bibinfo {pages} {4887} (\bibinfo {year} {2013})}\BibitemShut {NoStop}%
\bibitem [{\citenamefont {Hsu}\ \emph {et~al.}(2024)\citenamefont {Hsu}, \citenamefont {Tsai}, \citenamefont {Yeh},\ and\ \citenamefont {Yeh}}]{hsu_clarifying_2024}%
  \BibitemOpen
  \bibfield  {author} {\bibinfo {author} {\bibfnamefont {W.-L.}\ \bibnamefont {Hsu}}, \bibinfo {author} {\bibfnamefont {C.-W.}\ \bibnamefont {Tsai}}, \bibinfo {author} {\bibfnamefont {A.-C.}\ \bibnamefont {Yeh}},\ and\ \bibinfo {author} {\bibfnamefont {J.-W.}\ \bibnamefont {Yeh}},\ }\bibfield  {title} {\bibinfo {title} {Clarifying the four core effects of high-entropy materials},\ }\href {https://doi.org/10.1038/s41570-024-00602-5} {\bibfield  {journal} {\bibinfo  {journal} {Nature Reviews Chemistry}\ }\textbf {\bibinfo {volume} {8}},\ \bibinfo {pages} {471} (\bibinfo {year} {2024})}\BibitemShut {NoStop}%
\bibitem [{\citenamefont {Paul}\ \emph {et~al.}(2014)\citenamefont {Paul}, \citenamefont {Laurila}, \citenamefont {Vuorinen},\ and\ \citenamefont {Divinski}}]{paul_thermodynamics_2014}%
  \BibitemOpen
  \bibfield  {author} {\bibinfo {author} {\bibfnamefont {A.}~\bibnamefont {Paul}}, \bibinfo {author} {\bibfnamefont {T.}~\bibnamefont {Laurila}}, \bibinfo {author} {\bibfnamefont {V.}~\bibnamefont {Vuorinen}},\ and\ \bibinfo {author} {\bibfnamefont {S.~V.}\ \bibnamefont {Divinski}},\ }\href {https://doi.org/10.1007/978-3-319-07461-0} {\emph {\bibinfo {title} {Thermodynamics, {Diffusion} and the {Kirkendall} {Effect} in {Solids}}}}\ (\bibinfo  {publisher} {Springer International Publishing},\ \bibinfo {address} {Cham},\ \bibinfo {year} {2014})\BibitemShut {NoStop}%
\bibitem [{\citenamefont {DeHoff}\ and\ \citenamefont {Kulkarni}(2002)}]{dehoff_trouble_2002}%
  \BibitemOpen
  \bibfield  {author} {\bibinfo {author} {\bibfnamefont {R.~T.}\ \bibnamefont {DeHoff}}\ and\ \bibinfo {author} {\bibfnamefont {N.}~\bibnamefont {Kulkarni}},\ }\bibfield  {title} {\bibinfo {title} {The {Trouble} with {Diffusion}},\ }\href {https://doi.org/https://doi.org/10.1590/S1516-14392002000300002} {\bibfield  {journal} {\bibinfo  {journal} {Materials Research}\ }\textbf {\bibinfo {volume} {5}},\ \bibinfo {pages} {209} (\bibinfo {year} {2002})}\BibitemShut {NoStop}%
\bibitem [{\citenamefont {Dash}\ \emph {et~al.}(2022)\citenamefont {Dash}, \citenamefont {Paul}, \citenamefont {Sen}, \citenamefont {Divinski}, \citenamefont {Kundin}, \citenamefont {Steinbach}, \citenamefont {Grabowski},\ and\ \citenamefont {Zhang}}]{dash_recent_2022}%
  \BibitemOpen
  \bibfield  {author} {\bibinfo {author} {\bibfnamefont {A.}~\bibnamefont {Dash}}, \bibinfo {author} {\bibfnamefont {A.}~\bibnamefont {Paul}}, \bibinfo {author} {\bibfnamefont {S.}~\bibnamefont {Sen}}, \bibinfo {author} {\bibfnamefont {S.}~\bibnamefont {Divinski}}, \bibinfo {author} {\bibfnamefont {J.}~\bibnamefont {Kundin}}, \bibinfo {author} {\bibfnamefont {I.}~\bibnamefont {Steinbach}}, \bibinfo {author} {\bibfnamefont {B.}~\bibnamefont {Grabowski}},\ and\ \bibinfo {author} {\bibfnamefont {X.}~\bibnamefont {Zhang}},\ }\bibfield  {title} {\bibinfo {title} {Recent {Advances} in {Understanding} {Diffusion} in {Multiprincipal} {Element} {Systems}},\ }\href {https://doi.org/10.1146/annurev-matsci-081720-092213} {\bibfield  {journal} {\bibinfo  {journal} {Annual Review of Materials Research}\ }\textbf {\bibinfo {volume} {52}},\ \bibinfo {pages} {383} (\bibinfo {year} {2022})}\BibitemShut {NoStop}%
\bibitem [{\citenamefont {Xing}\ \emph {et~al.}(2024)\citenamefont {Xing}, \citenamefont {Rupert}, \citenamefont {Pan},\ and\ \citenamefont {Cao}}]{xing_neural_2024}%
  \BibitemOpen
  \bibfield  {author} {\bibinfo {author} {\bibfnamefont {B.}~\bibnamefont {Xing}}, \bibinfo {author} {\bibfnamefont {T.~J.}\ \bibnamefont {Rupert}}, \bibinfo {author} {\bibfnamefont {X.}~\bibnamefont {Pan}},\ and\ \bibinfo {author} {\bibfnamefont {P.}~\bibnamefont {Cao}},\ }\bibfield  {title} {\bibinfo {title} {Neural network kinetics for exploring diffusion multiplicity and chemical ordering in compositionally complex materials},\ }\href {https://doi.org/10.1038/s41467-024-47927-9} {\bibfield  {journal} {\bibinfo  {journal} {Nature Communications}\ }\textbf {\bibinfo {volume} {15}},\ \bibinfo {pages} {3879} (\bibinfo {year} {2024})}\BibitemShut {NoStop}%
\bibitem [{\citenamefont {Goiri}\ \emph {et~al.}(2019)\citenamefont {Goiri}, \citenamefont {Kolli},\ and\ \citenamefont {Van~der Ven}}]{goiri_role_2019}%
  \BibitemOpen
  \bibfield  {author} {\bibinfo {author} {\bibfnamefont {J.~G.}\ \bibnamefont {Goiri}}, \bibinfo {author} {\bibfnamefont {S.~K.}\ \bibnamefont {Kolli}},\ and\ \bibinfo {author} {\bibfnamefont {A.}~\bibnamefont {Van~der Ven}},\ }\bibfield  {title} {\bibinfo {title} {Role of short- and long-range ordering on diffusion in {Ni}-{Al} alloys},\ }\href {https://doi.org/10.1103/PhysRevMaterials.3.093402} {\bibfield  {journal} {\bibinfo  {journal} {Physical Review Materials}\ }\textbf {\bibinfo {volume} {3}},\ \bibinfo {pages} {093402} (\bibinfo {year} {2019})}\BibitemShut {NoStop}%
\bibitem [{\citenamefont {Zhang}\ \emph {et~al.}(2022{\natexlab{a}})\citenamefont {Zhang}, \citenamefont {Gadelmeier}, \citenamefont {Sen}, \citenamefont {Wang}, \citenamefont {Zhang}, \citenamefont {Zhong}, \citenamefont {Glatzel}, \citenamefont {Grabowski}, \citenamefont {Wilde},\ and\ \citenamefont {Divinski}}]{zhang_zr_2022}%
  \BibitemOpen
  \bibfield  {author} {\bibinfo {author} {\bibfnamefont {J.}~\bibnamefont {Zhang}}, \bibinfo {author} {\bibfnamefont {C.}~\bibnamefont {Gadelmeier}}, \bibinfo {author} {\bibfnamefont {S.}~\bibnamefont {Sen}}, \bibinfo {author} {\bibfnamefont {R.}~\bibnamefont {Wang}}, \bibinfo {author} {\bibfnamefont {X.}~\bibnamefont {Zhang}}, \bibinfo {author} {\bibfnamefont {Y.}~\bibnamefont {Zhong}}, \bibinfo {author} {\bibfnamefont {U.}~\bibnamefont {Glatzel}}, \bibinfo {author} {\bibfnamefont {B.}~\bibnamefont {Grabowski}}, \bibinfo {author} {\bibfnamefont {G.}~\bibnamefont {Wilde}},\ and\ \bibinfo {author} {\bibfnamefont {S.~V.}\ \bibnamefont {Divinski}},\ }\bibfield  {title} {\bibinfo {title} {Zr diffusion in {BCC} refractory high entropy alloys: {A} case of ‘non-sluggish’ diffusion behavior},\ }\href {https://doi.org/10.1016/j.actamat.2022.117970} {\bibfield  {journal} {\bibinfo  {journal} {Acta Materialia}\ }\textbf {\bibinfo {volume} {233}},\ \bibinfo {pages} {117970} (\bibinfo {year}
  {2022}{\natexlab{a}})}\BibitemShut {NoStop}%
\bibitem [{\citenamefont {Sen}\ \emph {et~al.}(2023)\citenamefont {Sen}, \citenamefont {Zhang}, \citenamefont {Rogal}, \citenamefont {Wilde}, \citenamefont {Grabowski},\ and\ \citenamefont {Divinski}}]{sen_anti-sluggish_2023}%
  \BibitemOpen
  \bibfield  {author} {\bibinfo {author} {\bibfnamefont {S.}~\bibnamefont {Sen}}, \bibinfo {author} {\bibfnamefont {X.}~\bibnamefont {Zhang}}, \bibinfo {author} {\bibfnamefont {L.}~\bibnamefont {Rogal}}, \bibinfo {author} {\bibfnamefont {G.}~\bibnamefont {Wilde}}, \bibinfo {author} {\bibfnamefont {B.}~\bibnamefont {Grabowski}},\ and\ \bibinfo {author} {\bibfnamefont {S.~V.}\ \bibnamefont {Divinski}},\ }\bibfield  {title} {\bibinfo {title} {‘{Anti}-sluggish’ {Ti} diffusion in {HCP} high-entropy alloys: {Chemical} complexity vs. lattice distortions},\ }\href {https://doi.org/10.1016/j.scriptamat.2022.115117} {\bibfield  {journal} {\bibinfo  {journal} {Scripta Materialia}\ }\textbf {\bibinfo {volume} {224}},\ \bibinfo {pages} {115117} (\bibinfo {year} {2023})}\BibitemShut {NoStop}%
\bibitem [{\citenamefont {Kirkaldy}\ and\ \citenamefont {Lane}(1966)}]{kirkaldy_diffusion_1966}%
  \BibitemOpen
  \bibfield  {author} {\bibinfo {author} {\bibfnamefont {J.~S.}\ \bibnamefont {Kirkaldy}}\ and\ \bibinfo {author} {\bibfnamefont {J.~E.}\ \bibnamefont {Lane}},\ }\bibfield  {title} {\bibinfo {title} {Diffusion in multicomponent metallic systems: ix. intrinsic diffusion behavior and the kirkendall effect in ternary substitutional solutions},\ }\href {https://doi.org/10.1139/p66-169} {\bibfield  {journal} {\bibinfo  {journal} {Canadian Journal of Physics}\ }\textbf {\bibinfo {volume} {44}},\ \bibinfo {pages} {2059} (\bibinfo {year} {1966})}\BibitemShut {NoStop}%
\bibitem [{\citenamefont {Esakkiraja}\ \emph {et~al.}(2019)\citenamefont {Esakkiraja}, \citenamefont {Pandey}, \citenamefont {Dash},\ and\ \citenamefont {Paul}}]{esakkiraja_pseudo-binary_2019}%
  \BibitemOpen
  \bibfield  {author} {\bibinfo {author} {\bibfnamefont {N.}~\bibnamefont {Esakkiraja}}, \bibinfo {author} {\bibfnamefont {K.}~\bibnamefont {Pandey}}, \bibinfo {author} {\bibfnamefont {A.}~\bibnamefont {Dash}},\ and\ \bibinfo {author} {\bibfnamefont {A.}~\bibnamefont {Paul}},\ }\bibfield  {title} {\bibinfo {title} {Pseudo-binary and pseudo-ternary diffusion couple methods for estimation of the diffusion coefficients in multicomponent systems and high entropy alloys},\ }\href {https://doi.org/10.1080/14786435.2019.1619027} {\bibfield  {journal} {\bibinfo  {journal} {Philosophical Magazine}\ }\textbf {\bibinfo {volume} {99}},\ \bibinfo {pages} {2236} (\bibinfo {year} {2019})}\BibitemShut {NoStop}%
\bibitem [{\citenamefont {Dash}\ \emph {et~al.}(2020)\citenamefont {Dash}, \citenamefont {Esakkiraja},\ and\ \citenamefont {Paul}}]{dash_solving_2020}%
  \BibitemOpen
  \bibfield  {author} {\bibinfo {author} {\bibfnamefont {A.}~\bibnamefont {Dash}}, \bibinfo {author} {\bibfnamefont {N.}~\bibnamefont {Esakkiraja}},\ and\ \bibinfo {author} {\bibfnamefont {A.}~\bibnamefont {Paul}},\ }\bibfield  {title} {\bibinfo {title} {Solving the issues of multicomponent diffusion in an equiatomic {NiCoFeCr} medium entropy alloy},\ }\href {https://doi.org/10.1016/j.actamat.2020.03.041} {\bibfield  {journal} {\bibinfo  {journal} {Acta Materialia}\ }\textbf {\bibinfo {volume} {193}},\ \bibinfo {pages} {163} (\bibinfo {year} {2020})}\BibitemShut {NoStop}%
\bibitem [{\citenamefont {Van~der Ven}\ \emph {et~al.}(2020)\citenamefont {Van~der Ven}, \citenamefont {Deng}, \citenamefont {Banerjee},\ and\ \citenamefont {Ong}}]{van_der_ven_rechargeable_2020}%
  \BibitemOpen
  \bibfield  {author} {\bibinfo {author} {\bibfnamefont {A.}~\bibnamefont {Van~der Ven}}, \bibinfo {author} {\bibfnamefont {Z.}~\bibnamefont {Deng}}, \bibinfo {author} {\bibfnamefont {S.}~\bibnamefont {Banerjee}},\ and\ \bibinfo {author} {\bibfnamefont {S.~P.}\ \bibnamefont {Ong}},\ }\bibfield  {title} {\bibinfo {title} {Rechargeable {Alkali}-{Ion} {Battery} {Materials}: {Theory} and {Computation}},\ }\href {https://doi.org/10.1021/acs.chemrev.9b00601} {\bibfield  {journal} {\bibinfo  {journal} {Chemical Reviews}\ }\textbf {\bibinfo {volume} {120}},\ \bibinfo {pages} {6977} (\bibinfo {year} {2020})}\BibitemShut {NoStop}%
\bibitem [{\citenamefont {Van~der Ven}\ \emph {et~al.}(2010)\citenamefont {Van~der Ven}, \citenamefont {Yu}, \citenamefont {Ceder},\ and\ \citenamefont {Thornton}}]{van_der_ven_vacancy_2010}%
  \BibitemOpen
  \bibfield  {author} {\bibinfo {author} {\bibfnamefont {A.}~\bibnamefont {Van~der Ven}}, \bibinfo {author} {\bibfnamefont {H.-C.}\ \bibnamefont {Yu}}, \bibinfo {author} {\bibfnamefont {G.}~\bibnamefont {Ceder}},\ and\ \bibinfo {author} {\bibfnamefont {K.}~\bibnamefont {Thornton}},\ }\bibfield  {title} {\bibinfo {title} {Vacancy mediated substitutional diffusion in binary crystalline solids},\ }\href {https://doi.org/10.1016/j.pmatsci.2009.08.001} {\bibfield  {journal} {\bibinfo  {journal} {Progress in Materials Science}\ }\textbf {\bibinfo {volume} {55}},\ \bibinfo {pages} {61} (\bibinfo {year} {2010})}\BibitemShut {NoStop}%
\bibitem [{\citenamefont {Balluffi}\ \emph {et~al.}(2005)\citenamefont {Balluffi}, \citenamefont {Allen},\ and\ \citenamefont {Carter}}]{balluffi_kinetics_2005}%
  \BibitemOpen
  \bibfield  {author} {\bibinfo {author} {\bibfnamefont {R.~W.}\ \bibnamefont {Balluffi}}, \bibinfo {author} {\bibfnamefont {S.~M.}\ \bibnamefont {Allen}},\ and\ \bibinfo {author} {\bibnamefont {Carter}},\ }\href {https://onlinelibrary.wiley.com/doi/abs/10.1002/0471749311.ch1} {\emph {\bibinfo {title} {Kinetics of {Materials}}}}\ (\bibinfo  {publisher} {John Wiley \& Sons, Ltd},\ \bibinfo {year} {2005})\BibitemShut {NoStop}%
\bibitem [{\citenamefont {Kirkaldy}\ \emph {et~al.}(1987)\citenamefont {Kirkaldy}, \citenamefont {Young},\ and\ \citenamefont {Lane}}]{kirkaldy_diffusion_1987}%
  \BibitemOpen
  \bibfield  {author} {\bibinfo {author} {\bibfnamefont {J.~S.}\ \bibnamefont {Kirkaldy}}, \bibinfo {author} {\bibfnamefont {D.~J.}\ \bibnamefont {Young}},\ and\ \bibinfo {author} {\bibfnamefont {J.~E.}\ \bibnamefont {Lane}},\ }\bibfield  {title} {\bibinfo {title} {Diffusion profiles associated with the onsager matrix in non-equilibrium {A}-{A}*-vacancy and {A}-{B}-vacancy solutions},\ }\href {https://doi.org/10.1016/0001-6160(87)90008-3} {\bibfield  {journal} {\bibinfo  {journal} {Acta Metallurgica}\ }\textbf {\bibinfo {volume} {35}},\ \bibinfo {pages} {1273} (\bibinfo {year} {1987})}\BibitemShut {NoStop}%
\bibitem [{\citenamefont {Shewmon}(1958)}]{shewmon_thermal_1958}%
  \BibitemOpen
  \bibfield  {author} {\bibinfo {author} {\bibfnamefont {P.}~\bibnamefont {Shewmon}},\ }\bibfield  {title} {\bibinfo {title} {Thermal {Diffusion} of {Vacancies} in {Zinc}},\ }\href {https://doi.org/10.1063/1.1744650} {\bibfield  {journal} {\bibinfo  {journal} {The Journal of Chemical Physics}\ }\textbf {\bibinfo {volume} {29}},\ \bibinfo {pages} {1032} (\bibinfo {year} {1958})}\BibitemShut {NoStop}%
\bibitem [{\citenamefont {Behara}\ and\ \citenamefont {Van~der Ven}(2024)}]{behara_role_2024}%
  \BibitemOpen
  \bibfield  {author} {\bibinfo {author} {\bibfnamefont {S.~S.}\ \bibnamefont {Behara}}\ and\ \bibinfo {author} {\bibfnamefont {A.}~\bibnamefont {Van~der Ven}},\ }\bibfield  {title} {\bibinfo {title} {Role of {Short}-{Range} {Order} on {Diffusion} {Coefficients} in the {Li}–{Mg} {Alloy}},\ }\href {https://doi.org/10.1021/acs.chemmater.4c02334} {\bibfield  {journal} {\bibinfo  {journal} {Chemistry of Materials}\ }\textbf {\bibinfo {volume} {36}},\ \bibinfo {pages} {11236} (\bibinfo {year} {2024})}\BibitemShut {NoStop}%
\bibitem [{\citenamefont {Deng}\ \emph {et~al.}(2022)\citenamefont {Deng}, \citenamefont {Mishra}, \citenamefont {Mahayoni}, \citenamefont {Ma}, \citenamefont {Tieu}, \citenamefont {Guillon}, \citenamefont {Chotard}, \citenamefont {Seznec}, \citenamefont {Cheetham}, \citenamefont {Masquelier}, \citenamefont {Gautam},\ and\ \citenamefont {Canepa}}]{deng_fundamental_2022}%
  \BibitemOpen
  \bibfield  {author} {\bibinfo {author} {\bibfnamefont {Z.}~\bibnamefont {Deng}}, \bibinfo {author} {\bibfnamefont {T.~P.}\ \bibnamefont {Mishra}}, \bibinfo {author} {\bibfnamefont {E.}~\bibnamefont {Mahayoni}}, \bibinfo {author} {\bibfnamefont {Q.}~\bibnamefont {Ma}}, \bibinfo {author} {\bibfnamefont {A.~J.~K.}\ \bibnamefont {Tieu}}, \bibinfo {author} {\bibfnamefont {O.}~\bibnamefont {Guillon}}, \bibinfo {author} {\bibfnamefont {J.-N.}\ \bibnamefont {Chotard}}, \bibinfo {author} {\bibfnamefont {V.}~\bibnamefont {Seznec}}, \bibinfo {author} {\bibfnamefont {A.~K.}\ \bibnamefont {Cheetham}}, \bibinfo {author} {\bibfnamefont {C.}~\bibnamefont {Masquelier}}, \bibinfo {author} {\bibfnamefont {G.~S.}\ \bibnamefont {Gautam}},\ and\ \bibinfo {author} {\bibfnamefont {P.}~\bibnamefont {Canepa}},\ }\bibfield  {title} {\bibinfo {title} {Fundamental investigations on the sodium-ion transport properties of mixed polyanion solid-state battery electrolytes},\ }\href {https://doi.org/10.1038/s41467-022-32190-7} {\bibfield
  {journal} {\bibinfo  {journal} {Nature Communications}\ }\textbf {\bibinfo {volume} {13}},\ \bibinfo {pages} {4470} (\bibinfo {year} {2022})}\BibitemShut {NoStop}%
\bibitem [{\citenamefont {Jadidi}\ \emph {et~al.}(2023)\citenamefont {Jadidi}, \citenamefont {Chen}, \citenamefont {Barroso-Luque},\ and\ \citenamefont {Ceder}}]{jadidi_kinetics_2023}%
  \BibitemOpen
  \bibfield  {author} {\bibinfo {author} {\bibfnamefont {Z.}~\bibnamefont {Jadidi}}, \bibinfo {author} {\bibfnamefont {T.}~\bibnamefont {Chen}}, \bibinfo {author} {\bibfnamefont {L.}~\bibnamefont {Barroso-Luque}},\ and\ \bibinfo {author} {\bibfnamefont {G.}~\bibnamefont {Ceder}},\ }\bibfield  {title} {\bibinfo {title} {Kinetics of {Li} {Transport} in {Vanadium}-{Based} {Disordered} {Rocksalt} {Structures}},\ }\href {https://doi.org/10.1021/acs.chemmater.3c01941} {\bibfield  {journal} {\bibinfo  {journal} {Chemistry of Materials}\ }\textbf {\bibinfo {volume} {35}},\ \bibinfo {pages} {9225} (\bibinfo {year} {2023})}\BibitemShut {NoStop}%
\bibitem [{\citenamefont {Canepa}(2023)}]{canepa_pushing_2023}%
  \BibitemOpen
  \bibfield  {author} {\bibinfo {author} {\bibfnamefont {P.}~\bibnamefont {Canepa}},\ }\bibfield  {title} {\bibinfo {title} {Pushing {Forward} {Simulation} {Techniques} of {Ion} {Transport} in {Ion} {Conductors} for {Energy} {Materials}},\ }\href {https://doi.org/10.1021/acsmaterialsau.2c00057} {\bibfield  {journal} {\bibinfo  {journal} {ACS Materials Au}\ }\textbf {\bibinfo {volume} {3}},\ \bibinfo {pages} {75} (\bibinfo {year} {2023})}\BibitemShut {NoStop}%
\bibitem [{\citenamefont {Cook}\ \emph {et~al.}(2024)\citenamefont {Cook}, \citenamefont {Kumar}, \citenamefont {Payne}, \citenamefont {Belcher}, \citenamefont {Borges}, \citenamefont {Wang}, \citenamefont {Walsh}, \citenamefont {Li}, \citenamefont {Devaraj}, \citenamefont {Zhang}, \citenamefont {Asta}, \citenamefont {Minor}, \citenamefont {Lavernia}, \citenamefont {Apelian},\ and\ \citenamefont {Ritchie}}]{cook_kink_2024}%
  \BibitemOpen
  \bibfield  {author} {\bibinfo {author} {\bibfnamefont {D.~H.}\ \bibnamefont {Cook}}, \bibinfo {author} {\bibfnamefont {P.}~\bibnamefont {Kumar}}, \bibinfo {author} {\bibfnamefont {M.~I.}\ \bibnamefont {Payne}}, \bibinfo {author} {\bibfnamefont {C.~H.}\ \bibnamefont {Belcher}}, \bibinfo {author} {\bibfnamefont {P.}~\bibnamefont {Borges}}, \bibinfo {author} {\bibfnamefont {W.}~\bibnamefont {Wang}}, \bibinfo {author} {\bibfnamefont {F.}~\bibnamefont {Walsh}}, \bibinfo {author} {\bibfnamefont {Z.}~\bibnamefont {Li}}, \bibinfo {author} {\bibfnamefont {A.}~\bibnamefont {Devaraj}}, \bibinfo {author} {\bibfnamefont {M.}~\bibnamefont {Zhang}}, \bibinfo {author} {\bibfnamefont {M.}~\bibnamefont {Asta}}, \bibinfo {author} {\bibfnamefont {A.~M.}\ \bibnamefont {Minor}}, \bibinfo {author} {\bibfnamefont {E.~J.}\ \bibnamefont {Lavernia}}, \bibinfo {author} {\bibfnamefont {D.}~\bibnamefont {Apelian}},\ and\ \bibinfo {author} {\bibfnamefont {R.~O.}\ \bibnamefont {Ritchie}},\ }\bibfield  {title} {\bibinfo {title} {Kink bands
  promote exceptional fracture resistance in a {NbTaTiHf} refractory medium-entropy alloy},\ }\href {https://doi.org/10.1126/science.adn2428} {\bibfield  {journal} {\bibinfo  {journal} {Science}\ }\textbf {\bibinfo {volume} {384}},\ \bibinfo {pages} {178} (\bibinfo {year} {2024})}\BibitemShut {NoStop}%
\bibitem [{\citenamefont {Kumar}\ \emph {et~al.}(2024)\citenamefont {Kumar}, \citenamefont {Gou}, \citenamefont {Cook}, \citenamefont {Payne}, \citenamefont {Morrison}, \citenamefont {Wang}, \citenamefont {Zhang}, \citenamefont {Asta}, \citenamefont {Minor}, \citenamefont {Cao}, \citenamefont {Li},\ and\ \citenamefont {Ritchie}}]{kumar_degradation_2024}%
  \BibitemOpen
  \bibfield  {author} {\bibinfo {author} {\bibfnamefont {P.}~\bibnamefont {Kumar}}, \bibinfo {author} {\bibfnamefont {X.}~\bibnamefont {Gou}}, \bibinfo {author} {\bibfnamefont {D.~H.}\ \bibnamefont {Cook}}, \bibinfo {author} {\bibfnamefont {M.~I.}\ \bibnamefont {Payne}}, \bibinfo {author} {\bibfnamefont {N.~J.}\ \bibnamefont {Morrison}}, \bibinfo {author} {\bibfnamefont {W.}~\bibnamefont {Wang}}, \bibinfo {author} {\bibfnamefont {M.}~\bibnamefont {Zhang}}, \bibinfo {author} {\bibfnamefont {M.}~\bibnamefont {Asta}}, \bibinfo {author} {\bibfnamefont {A.~M.}\ \bibnamefont {Minor}}, \bibinfo {author} {\bibfnamefont {R.}~\bibnamefont {Cao}}, \bibinfo {author} {\bibfnamefont {Y.}~\bibnamefont {Li}},\ and\ \bibinfo {author} {\bibfnamefont {R.~O.}\ \bibnamefont {Ritchie}},\ }\bibfield  {title} {\bibinfo {title} {Degradation of the mechanical properties of {NbMoTaW} refractory high-entropy alloy in tension},\ }\href {https://doi.org/10.1016/j.actamat.2024.120297} {\bibfield  {journal} {\bibinfo  {journal} {Acta
  Materialia}\ }\textbf {\bibinfo {volume} {279}},\ \bibinfo {pages} {120297} (\bibinfo {year} {2024})}\BibitemShut {NoStop}%
\bibitem [{\citenamefont {George}\ \emph {et~al.}(2019)\citenamefont {George}, \citenamefont {Raabe},\ and\ \citenamefont {Ritchie}}]{george_high-entropy_2019}%
  \BibitemOpen
  \bibfield  {author} {\bibinfo {author} {\bibfnamefont {E.~P.}\ \bibnamefont {George}}, \bibinfo {author} {\bibfnamefont {D.}~\bibnamefont {Raabe}},\ and\ \bibinfo {author} {\bibfnamefont {R.~O.}\ \bibnamefont {Ritchie}},\ }\bibfield  {title} {\bibinfo {title} {High-entropy alloys},\ }\href {https://doi.org/10.1038/s41578-019-0121-4} {\bibfield  {journal} {\bibinfo  {journal} {Nature Reviews Materials}\ }\textbf {\bibinfo {volume} {4}},\ \bibinfo {pages} {515} (\bibinfo {year} {2019})}\BibitemShut {NoStop}%
\bibitem [{\citenamefont {Mak}\ \emph {et~al.}(2021)\citenamefont {Mak}, \citenamefont {Yin},\ and\ \citenamefont {Curtin}}]{mak_ductility_2021}%
  \BibitemOpen
  \bibfield  {author} {\bibinfo {author} {\bibfnamefont {E.}~\bibnamefont {Mak}}, \bibinfo {author} {\bibfnamefont {B.}~\bibnamefont {Yin}},\ and\ \bibinfo {author} {\bibfnamefont {W.~A.}\ \bibnamefont {Curtin}},\ }\bibfield  {title} {\bibinfo {title} {A ductility criterion for bcc high entropy alloys},\ }\href {https://doi.org/10.1016/j.jmps.2021.104389} {\bibfield  {journal} {\bibinfo  {journal} {Journal of the Mechanics and Physics of Solids}\ }\textbf {\bibinfo {volume} {152}},\ \bibinfo {pages} {104389} (\bibinfo {year} {2021})}\BibitemShut {NoStop}%
\bibitem [{\citenamefont {Li}\ \emph {et~al.}(2020)\citenamefont {Li}, \citenamefont {Chen}, \citenamefont {Zheng}, \citenamefont {Zuo},\ and\ \citenamefont {Ong}}]{li_complex_2020}%
  \BibitemOpen
  \bibfield  {author} {\bibinfo {author} {\bibfnamefont {X.-G.}\ \bibnamefont {Li}}, \bibinfo {author} {\bibfnamefont {C.}~\bibnamefont {Chen}}, \bibinfo {author} {\bibfnamefont {H.}~\bibnamefont {Zheng}}, \bibinfo {author} {\bibfnamefont {Y.}~\bibnamefont {Zuo}},\ and\ \bibinfo {author} {\bibfnamefont {S.~P.}\ \bibnamefont {Ong}},\ }\bibfield  {title} {\bibinfo {title} {Complex strengthening mechanisms in the {NbMoTaW} multi-principal element alloy},\ }\href {https://doi.org/10.1038/s41524-020-0339-0} {\bibfield  {journal} {\bibinfo  {journal} {npj Computational Materials}\ }\textbf {\bibinfo {volume} {6}},\ \bibinfo {pages} {70} (\bibinfo {year} {2020})}\BibitemShut {NoStop}%
\bibitem [{\citenamefont {Kehr}\ \emph {et~al.}(1989)\citenamefont {Kehr}, \citenamefont {Binder},\ and\ \citenamefont {Reulein}}]{kehr_mobility_1989}%
  \BibitemOpen
  \bibfield  {author} {\bibinfo {author} {\bibfnamefont {K.~W.}\ \bibnamefont {Kehr}}, \bibinfo {author} {\bibfnamefont {K.}~\bibnamefont {Binder}},\ and\ \bibinfo {author} {\bibfnamefont {S.~M.}\ \bibnamefont {Reulein}},\ }\bibfield  {title} {\bibinfo {title} {Mobility, interdiffusion, and tracer diffusion in lattice-gas models of two-component alloys},\ }\href {https://doi.org/10.1103/PhysRevB.39.4891} {\bibfield  {journal} {\bibinfo  {journal} {Physical Review B}\ }\textbf {\bibinfo {volume} {39}},\ \bibinfo {pages} {4891} (\bibinfo {year} {1989})}\BibitemShut {NoStop}%
\bibitem [{\citenamefont {Green}(1954)}]{green_markoff_1954}%
  \BibitemOpen
  \bibfield  {author} {\bibinfo {author} {\bibfnamefont {M.~S.}\ \bibnamefont {Green}},\ }\bibfield  {title} {\bibinfo {title} {Markoff {Random} {Processes} and the {Statistical} {Mechanics} of {Time}‐{Dependent} {Phenomena}. {II}. {Irreversible} {Processes} in {Fluids}},\ }\href {https://doi.org/10.1063/1.1740082} {\bibfield  {journal} {\bibinfo  {journal} {The Journal of Chemical Physics}\ }\textbf {\bibinfo {volume} {22}},\ \bibinfo {pages} {398} (\bibinfo {year} {1954})}\BibitemShut {NoStop}%
\bibitem [{\citenamefont {Kubo}(1957)}]{kubo_statistical-mechanical_1957}%
  \BibitemOpen
  \bibfield  {author} {\bibinfo {author} {\bibfnamefont {R.}~\bibnamefont {Kubo}},\ }\bibfield  {title} {\bibinfo {title} {Statistical-{Mechanical} {Theory} of {Irreversible} {Processes}. {I}. {General} {Theory} and {Simple} {Applications} to {Magnetic} and {Conduction} {Problems}},\ }\href {https://doi.org/10.1143/JPSJ.12.570} {\bibfield  {journal} {\bibinfo  {journal} {Journal of the Physical Society of Japan}\ }\textbf {\bibinfo {volume} {12}},\ \bibinfo {pages} {570} (\bibinfo {year} {1957})}\BibitemShut {NoStop}%
\bibitem [{\citenamefont {Vineyard}(1957)}]{vineyard_frequency_1957}%
  \BibitemOpen
  \bibfield  {author} {\bibinfo {author} {\bibfnamefont {G.~H.}\ \bibnamefont {Vineyard}},\ }\bibfield  {title} {\bibinfo {title} {Frequency factors and isotope effects in solid state rate processes},\ }\href {https://doi.org/10.1016/0022-3697(57)90059-8} {\bibfield  {journal} {\bibinfo  {journal} {Journal of Physics and Chemistry of Solids}\ }\textbf {\bibinfo {volume} {3}},\ \bibinfo {pages} {121} (\bibinfo {year} {1957})}\BibitemShut {NoStop}%
\bibitem [{\citenamefont {Binder}\ and\ \citenamefont {Heermann}(2010)}]{binder_monte_2010}%
  \BibitemOpen
  \bibfield  {author} {\bibinfo {author} {\bibfnamefont {K.}~\bibnamefont {Binder}}\ and\ \bibinfo {author} {\bibfnamefont {D.~W.}\ \bibnamefont {Heermann}},\ }\href {https://doi.org/10.1007/978-3-642-03163-2} {\emph {\bibinfo {title} {Monte {Carlo} {Simulation} in {Statistical} {Physics}: {An} {Introduction}}}},\ \bibinfo {series} {Graduate {Texts} in {Physics}}, Vol.~\bibinfo {volume} {0}\ (\bibinfo  {publisher} {Springer},\ \bibinfo {address} {Berlin, Heidelberg},\ \bibinfo {year} {2010})\BibitemShut {NoStop}%
\bibitem [{\citenamefont {Sanchez}\ \emph {et~al.}(1984)\citenamefont {Sanchez}, \citenamefont {Ducastelle},\ and\ \citenamefont {Gratias}}]{sanchez_generalized_1984}%
  \BibitemOpen
  \bibfield  {author} {\bibinfo {author} {\bibfnamefont {J.~M.}\ \bibnamefont {Sanchez}}, \bibinfo {author} {\bibfnamefont {F.}~\bibnamefont {Ducastelle}},\ and\ \bibinfo {author} {\bibfnamefont {D.}~\bibnamefont {Gratias}},\ }\bibfield  {title} {\bibinfo {title} {Generalized cluster description of multicomponent systems},\ }\href {https://doi.org/10.1016/0378-4371(84)90096-7} {\bibfield  {journal} {\bibinfo  {journal} {Physica A: Statistical Mechanics and its Applications}\ }\textbf {\bibinfo {volume} {128}},\ \bibinfo {pages} {334} (\bibinfo {year} {1984})}\BibitemShut {NoStop}%
\bibitem [{\citenamefont {Ozolins}(1996)}]{ozolins_notitle_1996}%
  \BibitemOpen
  \bibfield  {author} {\bibinfo {author} {\bibfnamefont {V.}~\bibnamefont {Ozolins}},\ }\href@noop {} {Ph.D. thesis},\ \bibinfo  {school} {Royal Institute of Technology}, \bibinfo {address} {Stockholm, Sweden} (\bibinfo {year} {1996})\BibitemShut {NoStop}%
\bibitem [{\citenamefont {Morgan}\ \emph {et~al.}(1998)\citenamefont {Morgan}, \citenamefont {Althoff},\ and\ \citenamefont {de~Fontaine}}]{morgan_local_1998}%
  \BibitemOpen
  \bibfield  {author} {\bibinfo {author} {\bibfnamefont {D.}~\bibnamefont {Morgan}}, \bibinfo {author} {\bibfnamefont {J.~D.}\ \bibnamefont {Althoff}},\ and\ \bibinfo {author} {\bibfnamefont {D.}~\bibnamefont {de~Fontaine}},\ }\bibfield  {title} {\bibinfo {title} {Local environment effects in the vibrational properties of disordered alloys: {An} embedded-atom method study of {Ni3Al} and {Cu3Au}},\ }\href {https://doi.org/10.1361/105497198770341752} {\bibfield  {journal} {\bibinfo  {journal} {Journal of Phase Equilibria}\ }\textbf {\bibinfo {volume} {19}},\ \bibinfo {pages} {559} (\bibinfo {year} {1998})}\BibitemShut {NoStop}%
\bibitem [{\citenamefont {Natarajan}\ and\ \citenamefont {Van~der Ven}(2020)}]{natarajan_linking_2020}%
  \BibitemOpen
  \bibfield  {author} {\bibinfo {author} {\bibfnamefont {A.~R.}\ \bibnamefont {Natarajan}}\ and\ \bibinfo {author} {\bibfnamefont {A.}~\bibnamefont {Van~der Ven}},\ }\bibfield  {title} {\bibinfo {title} {Linking electronic structure calculations to generalized stacking fault energies in multicomponent alloys},\ }\href {https://doi.org/10.1038/s41524-020-0348-z} {\bibfield  {journal} {\bibinfo  {journal} {npj Computational Materials}\ }\textbf {\bibinfo {volume} {6}},\ \bibinfo {pages} {80} (\bibinfo {year} {2020})}\BibitemShut {NoStop}%
\bibitem [{\citenamefont {Van~der Ven}\ \emph {et~al.}(2001)\citenamefont {Van~der Ven}, \citenamefont {Ceder}, \citenamefont {Asta},\ and\ \citenamefont {Tepesch}}]{van_der_ven_first-principles_2001}%
  \BibitemOpen
  \bibfield  {author} {\bibinfo {author} {\bibfnamefont {A.}~\bibnamefont {Van~der Ven}}, \bibinfo {author} {\bibfnamefont {G.}~\bibnamefont {Ceder}}, \bibinfo {author} {\bibfnamefont {M.}~\bibnamefont {Asta}},\ and\ \bibinfo {author} {\bibfnamefont {P.~D.}\ \bibnamefont {Tepesch}},\ }\bibfield  {title} {\bibinfo {title} {First-principles theory of ionic diffusion with nondilute carriers},\ }\href {https://doi.org/10.1103/PhysRevB.64.184307} {\bibfield  {journal} {\bibinfo  {journal} {Physical Review B}\ }\textbf {\bibinfo {volume} {64}},\ \bibinfo {pages} {184307} (\bibinfo {year} {2001})}\BibitemShut {NoStop}%
\bibitem [{\citenamefont {Zhang}\ \emph {et~al.}(2022{\natexlab{b}})\citenamefont {Zhang}, \citenamefont {Divinski},\ and\ \citenamefont {Grabowski}}]{zhang_ab_2022}%
  \BibitemOpen
  \bibfield  {author} {\bibinfo {author} {\bibfnamefont {X.}~\bibnamefont {Zhang}}, \bibinfo {author} {\bibfnamefont {S.~V.}\ \bibnamefont {Divinski}},\ and\ \bibinfo {author} {\bibfnamefont {B.}~\bibnamefont {Grabowski}},\ }\bibfield  {title} {\bibinfo {title} {\textit{{Ab} initio} prediction of vacancy energetics in {HCP} {Al}-{Hf}-{Sc}-{Ti}-{Zr} high entropy alloys and the subsystems},\ }\href {https://doi.org/10.1016/j.actamat.2022.117677} {\bibfield  {journal} {\bibinfo  {journal} {Acta Materialia}\ }\textbf {\bibinfo {volume} {227}},\ \bibinfo {pages} {117677} (\bibinfo {year} {2022}{\natexlab{b}})}\BibitemShut {NoStop}%
\bibitem [{\citenamefont {Barroso-Luque}\ and\ \citenamefont {Ceder}(2024)}]{barroso-luque_cluster_2024}%
  \BibitemOpen
  \bibfield  {author} {\bibinfo {author} {\bibfnamefont {L.}~\bibnamefont {Barroso-Luque}}\ and\ \bibinfo {author} {\bibfnamefont {G.}~\bibnamefont {Ceder}},\ }\bibfield  {title} {\bibinfo {title} {The cluster decomposition of the configurational energy of multicomponent alloys},\ }\href {https://doi.org/10.1038/s41524-024-01338-y} {\bibfield  {journal} {\bibinfo  {journal} {npj Computational Materials}\ }\textbf {\bibinfo {volume} {10}},\ \bibinfo {pages} {158} (\bibinfo {year} {2024})}\BibitemShut {NoStop}%
\bibitem [{\citenamefont {Müller}\ and\ \citenamefont {Natarajan}(2025)}]{muller_constructing_2025}%
  \BibitemOpen
  \bibfield  {author} {\bibinfo {author} {\bibfnamefont {Y.~L.}\ \bibnamefont {Müller}}\ and\ \bibinfo {author} {\bibfnamefont {A.~R.}\ \bibnamefont {Natarajan}},\ }\bibfield  {title} {\bibinfo {title} {Constructing multicomponent cluster expansions with machine-learning and chemical embedding},\ }\href {https://doi.org/10.1038/s41524-025-01543-3} {\bibfield  {journal} {\bibinfo  {journal} {npj Computational Materials}\ }\textbf {\bibinfo {volume} {11}},\ \bibinfo {pages} {1} (\bibinfo {year} {2025})}\BibitemShut {NoStop}%
\bibitem [{\citenamefont {Batatia}\ \emph {et~al.}(2023)\citenamefont {Batatia}, \citenamefont {Kovács}, \citenamefont {Simm}, \citenamefont {Ortner},\ and\ \citenamefont {Csányi}}]{batatia_mace_2023}%
  \BibitemOpen
  \bibfield  {author} {\bibinfo {author} {\bibfnamefont {I.}~\bibnamefont {Batatia}}, \bibinfo {author} {\bibfnamefont {D.~P.}\ \bibnamefont {Kovács}}, \bibinfo {author} {\bibfnamefont {G.~N.~C.}\ \bibnamefont {Simm}}, \bibinfo {author} {\bibfnamefont {C.}~\bibnamefont {Ortner}},\ and\ \bibinfo {author} {\bibfnamefont {G.}~\bibnamefont {Csányi}},\ }\href {https://doi.org/10.48550/arXiv.2206.07697} {\bibinfo {title} {{MACE}: {Higher} {Order} {Equivariant} {Message} {Passing} {Neural} {Networks} for {Fast} and {Accurate} {Force} {Fields}}} (\bibinfo {year} {2023}),\ \bibinfo {note} {arXiv:2206.07697 [stat]}\BibitemShut {NoStop}%
\bibitem [{\citenamefont {Miracle}\ \emph {et~al.}(2024)\citenamefont {Miracle}, \citenamefont {Senkov}, \citenamefont {Frey}, \citenamefont {Rao},\ and\ \citenamefont {Pollock}}]{miracle_strength_2024}%
  \BibitemOpen
  \bibfield  {author} {\bibinfo {author} {\bibfnamefont {D.~B.}\ \bibnamefont {Miracle}}, \bibinfo {author} {\bibfnamefont {O.~N.}\ \bibnamefont {Senkov}}, \bibinfo {author} {\bibfnamefont {C.}~\bibnamefont {Frey}}, \bibinfo {author} {\bibfnamefont {S.}~\bibnamefont {Rao}},\ and\ \bibinfo {author} {\bibfnamefont {T.~M.}\ \bibnamefont {Pollock}},\ }\bibfield  {title} {\bibinfo {title} {Strength \textit{vs} temperature for refractory complex concentrated alloys ({RCCAs}): {A} critical comparison with refractory {BCC} elements and dilute alloys},\ }\href {https://doi.org/10.1016/j.actamat.2024.119692} {\bibfield  {journal} {\bibinfo  {journal} {Acta Materialia}\ }\textbf {\bibinfo {volume} {266}},\ \bibinfo {pages} {119692} (\bibinfo {year} {2024})}\BibitemShut {NoStop}%
\bibitem [{\citenamefont {Natarajan}\ \emph {et~al.}(2020)\citenamefont {Natarajan}, \citenamefont {Dolin},\ and\ \citenamefont {Van~der Ven}}]{natarajan_crystallography_2020}%
  \BibitemOpen
  \bibfield  {author} {\bibinfo {author} {\bibfnamefont {A.~R.}\ \bibnamefont {Natarajan}}, \bibinfo {author} {\bibfnamefont {P.}~\bibnamefont {Dolin}},\ and\ \bibinfo {author} {\bibfnamefont {A.}~\bibnamefont {Van~der Ven}},\ }\bibfield  {title} {\bibinfo {title} {Crystallography, thermodynamics and phase transitions in refractory binary alloys},\ }\href {https://doi.org/10.1016/j.actamat.2020.08.034} {\bibfield  {journal} {\bibinfo  {journal} {Acta Materialia}\ }\textbf {\bibinfo {volume} {200}},\ \bibinfo {pages} {171} (\bibinfo {year} {2020})}\BibitemShut {NoStop}%
\bibitem [{\citenamefont {Lee}\ \emph {et~al.}(2026{\natexlab{a}})\citenamefont {Lee}, \citenamefont {Müller},\ and\ \citenamefont {Natarajan}}]{lee_modeling_2026}%
  \BibitemOpen
  \bibfield  {author} {\bibinfo {author} {\bibfnamefont {D.~K.~J.}\ \bibnamefont {Lee}}, \bibinfo {author} {\bibfnamefont {Y.~L.}\ \bibnamefont {Müller}},\ and\ \bibinfo {author} {\bibfnamefont {A.~R.}\ \bibnamefont {Natarajan}},\ }\bibfield  {title} {\bibinfo {title} {Modeling the equilibrium vacancy concentration in multi-principal element alloys from first-principles},\ }\href {https://doi.org/10.1016/j.actamat.2025.121752} {\bibfield  {journal} {\bibinfo  {journal} {Acta Materialia}\ }\textbf {\bibinfo {volume} {304}},\ \bibinfo {pages} {121752} (\bibinfo {year} {2026}{\natexlab{a}})}\BibitemShut {NoStop}%
\bibitem [{\citenamefont {Lee}\ \emph {et~al.}(2026{\natexlab{b}})\citenamefont {Lee}, \citenamefont {Saxena}, \citenamefont {Van Der~Ven},\ and\ \citenamefont {Natarajan}}]{lee_first-principles_2026}%
  \BibitemOpen
  \bibfield  {author} {\bibinfo {author} {\bibfnamefont {D.~K.~J.}\ \bibnamefont {Lee}}, \bibinfo {author} {\bibfnamefont {S.}~\bibnamefont {Saxena}}, \bibinfo {author} {\bibfnamefont {A.}~\bibnamefont {Van Der~Ven}},\ and\ \bibinfo {author} {\bibfnamefont {A.~R.}\ \bibnamefont {Natarajan}},\ }\bibfield  {title} {\bibinfo {title} {First-principles calculations of the thermodynamic factor of multicomponent alloys},\ }\href@noop {} {\bibfield  {journal} {\bibinfo  {journal} {Under Preparation}\ } (\bibinfo {year} {2026}{\natexlab{b}})}\BibitemShut {NoStop}%
\bibitem [{\citenamefont {Darken}(1951)}]{darken_formal_1951}%
  \BibitemOpen
  \bibfield  {author} {\bibinfo {author} {\bibfnamefont {L.~S.}\ \bibnamefont {Darken}},\ }\bibfield  {title} {\bibinfo {title} {Formal basis of diffusion theory},\ }\href@noop {} {\bibfield  {journal} {\bibinfo  {journal} {Atom movements}\ ,\ \bibinfo {pages} {1}} (\bibinfo {year} {1951})}\BibitemShut {NoStop}%
\bibitem [{\citenamefont {Yeh}(2006)}]{yeh_recent_2006}%
  \BibitemOpen
  \bibfield  {author} {\bibinfo {author} {\bibfnamefont {J.~W.}\ \bibnamefont {Yeh}},\ }\bibfield  {title} {\bibinfo {title} {Recent progress in high-entropy alloys},\ }\href {https://inis.iaea.org/records/7f3gk-s8e75} {\bibfield  {journal} {\bibinfo  {journal} {Annales De Chimie}\ }\textbf {\bibinfo {volume} {31}},\ \bibinfo {pages} {633} (\bibinfo {year} {2006})}\BibitemShut {NoStop}%
\bibitem [{\citenamefont {Yeh}\ \emph {et~al.}(2007)\citenamefont {Yeh}, \citenamefont {Chen}, \citenamefont {Lin},\ and\ \citenamefont {Chen}}]{yeh_high-entropy_2007}%
  \BibitemOpen
  \bibfield  {author} {\bibinfo {author} {\bibfnamefont {J.~W.}\ \bibnamefont {Yeh}}, \bibinfo {author} {\bibfnamefont {Y.~L.}\ \bibnamefont {Chen}}, \bibinfo {author} {\bibfnamefont {S.~J.}\ \bibnamefont {Lin}},\ and\ \bibinfo {author} {\bibfnamefont {S.~K.}\ \bibnamefont {Chen}},\ }\bibfield  {title} {\bibinfo {title} {High-{Entropy} {Alloys} – {A} {New} {Era} of {Exploitation}},\ }\href {https://doi.org/10.4028/www.scientific.net/MSF.560.1} {\bibfield  {journal} {\bibinfo  {journal} {Materials Science Forum}\ }\textbf {\bibinfo {volume} {560}},\ \bibinfo {pages} {1} (\bibinfo {year} {2007})}\BibitemShut {NoStop}%
\bibitem [{\citenamefont {Miracle}(2017)}]{miracle_high-entropy_2017}%
  \BibitemOpen
  \bibfield  {author} {\bibinfo {author} {\bibfnamefont {D.~B.}\ \bibnamefont {Miracle}},\ }\bibfield  {title} {\bibinfo {title} {High-{Entropy} {Alloys}: {A} {Current} {Evaluation} of {Founding} {Ideas} and {Core} {Effects} and {Exploring} “{Nonlinear} {Alloys}”},\ }\href {https://doi.org/10.1007/s11837-017-2527-z} {\bibfield  {journal} {\bibinfo  {journal} {JOM}\ }\textbf {\bibinfo {volume} {69}},\ \bibinfo {pages} {2130} (\bibinfo {year} {2017})}\BibitemShut {NoStop}%
\bibitem [{\citenamefont {Daw}\ and\ \citenamefont {Chandross}(2021)}]{daw_sluggish_2021}%
  \BibitemOpen
  \bibfield  {author} {\bibinfo {author} {\bibfnamefont {M.~S.}\ \bibnamefont {Daw}}\ and\ \bibinfo {author} {\bibfnamefont {M.}~\bibnamefont {Chandross}},\ }\bibfield  {title} {\bibinfo {title} {Sluggish diffusion in random equimolar {FCC} alloys},\ }\href {https://doi.org/10.1103/PhysRevMaterials.5.043603} {\bibfield  {journal} {\bibinfo  {journal} {Physical Review Materials}\ }\textbf {\bibinfo {volume} {5}},\ \bibinfo {pages} {043603} (\bibinfo {year} {2021})}\BibitemShut {NoStop}%
\bibitem [{\citenamefont {Moleko}\ \emph {et~al.}(1989)\citenamefont {Moleko}, \citenamefont {Allnatt},\ and\ \citenamefont {Allnatt}}]{moleko_self-consistent_1989}%
  \BibitemOpen
  \bibfield  {author} {\bibinfo {author} {\bibfnamefont {L.~K.}\ \bibnamefont {Moleko}}, \bibinfo {author} {\bibfnamefont {A.~R.}\ \bibnamefont {Allnatt}},\ and\ \bibinfo {author} {\bibfnamefont {E.~L.}\ \bibnamefont {Allnatt}},\ }\bibfield  {title} {\bibinfo {title} {A self-consistent theory of matter transport in a random lattice gas and some simulation results},\ }\href {https://doi.org/10.1080/01418618908220335} {\bibfield  {journal} {\bibinfo  {journal} {Philosophical Magazine A}\ }\textbf {\bibinfo {volume} {59}},\ \bibinfo {pages} {141} (\bibinfo {year} {1989})}\BibitemShut {NoStop}%
\bibitem [{\citenamefont {Manning}(1971)}]{manning_correlation_1971}%
  \BibitemOpen
  \bibfield  {author} {\bibinfo {author} {\bibfnamefont {J.~R.}\ \bibnamefont {Manning}},\ }\bibfield  {title} {\bibinfo {title} {Correlation {Factors} for {Diffusion} in {Nondilute} {Alloys}},\ }\href {https://doi.org/10.1103/PhysRevB.4.1111} {\bibfield  {journal} {\bibinfo  {journal} {Physical Review B}\ }\textbf {\bibinfo {volume} {4}},\ \bibinfo {pages} {1111} (\bibinfo {year} {1971})}\BibitemShut {NoStop}%
\bibitem [{\citenamefont {Mishin}\ and\ \citenamefont {Farkas}(1997)}]{mishin_monte_1997}%
  \BibitemOpen
  \bibfield  {author} {\bibinfo {author} {\bibfnamefont {Y.}~\bibnamefont {Mishin}}\ and\ \bibinfo {author} {\bibfnamefont {D.}~\bibnamefont {Farkas}},\ }\bibfield  {title} {\bibinfo {title} {Monte {Carlo} simulation of correlation effects in a random bcc alloy},\ }\href {https://doi.org/10.1080/01418619708210291} {\bibfield  {journal} {\bibinfo  {journal} {Philosophical Magazine A}\ }\textbf {\bibinfo {volume} {75}},\ \bibinfo {pages} {201} (\bibinfo {year} {1997})}\BibitemShut {NoStop}%
\bibitem [{\citenamefont {Sykes}\ and\ \citenamefont {Essam}(1964)}]{sykes_critical_1964}%
  \BibitemOpen
  \bibfield  {author} {\bibinfo {author} {\bibfnamefont {M.~F.}\ \bibnamefont {Sykes}}\ and\ \bibinfo {author} {\bibfnamefont {J.~W.}\ \bibnamefont {Essam}},\ }\bibfield  {title} {\bibinfo {title} {Critical {Percolation} {Probabilities} by {Series} {Methods}},\ }\href {https://doi.org/10.1103/PhysRev.133.A310} {\bibfield  {journal} {\bibinfo  {journal} {Physical Review}\ }\textbf {\bibinfo {volume} {133}},\ \bibinfo {pages} {A310} (\bibinfo {year} {1964})}\BibitemShut {NoStop}%
\bibitem [{\citenamefont {Zhang}\ \emph {et~al.}(2025)\citenamefont {Zhang}, \citenamefont {Divinski},\ and\ \citenamefont {Grabowski}}]{zhang_ab_2025}%
  \BibitemOpen
  \bibfield  {author} {\bibinfo {author} {\bibfnamefont {X.}~\bibnamefont {Zhang}}, \bibinfo {author} {\bibfnamefont {S.~V.}\ \bibnamefont {Divinski}},\ and\ \bibinfo {author} {\bibfnamefont {B.}~\bibnamefont {Grabowski}},\ }\bibfield  {title} {\bibinfo {title} {Ab initio machine-learning unveils strong anharmonicity in non-{Arrhenius} self-diffusion of tungsten},\ }\href {https://doi.org/10.1038/s41467-024-55759-w} {\bibfield  {journal} {\bibinfo  {journal} {Nature Communications}\ }\textbf {\bibinfo {volume} {16}},\ \bibinfo {pages} {394} (\bibinfo {year} {2025})}\BibitemShut {NoStop}%
\bibitem [{\citenamefont {Sangiovanni}\ \emph {et~al.}(2019)\citenamefont {Sangiovanni}, \citenamefont {Klarbring}, \citenamefont {Smirnova}, \citenamefont {Skripnyak}, \citenamefont {Gambino}, \citenamefont {Mrovec}, \citenamefont {Simak},\ and\ \citenamefont {Abrikosov}}]{sangiovanni_superioniclike_2019}%
  \BibitemOpen
  \bibfield  {author} {\bibinfo {author} {\bibfnamefont {D.}~\bibnamefont {Sangiovanni}}, \bibinfo {author} {\bibfnamefont {J.}~\bibnamefont {Klarbring}}, \bibinfo {author} {\bibfnamefont {D.}~\bibnamefont {Smirnova}}, \bibinfo {author} {\bibfnamefont {N.}~\bibnamefont {Skripnyak}}, \bibinfo {author} {\bibfnamefont {D.}~\bibnamefont {Gambino}}, \bibinfo {author} {\bibfnamefont {M.}~\bibnamefont {Mrovec}}, \bibinfo {author} {\bibfnamefont {S.}~\bibnamefont {Simak}},\ and\ \bibinfo {author} {\bibfnamefont {I.}~\bibnamefont {Abrikosov}},\ }\bibfield  {title} {\bibinfo {title} {Superioniclike {Diffusion} in an {Elemental} {Crystal}: bcc {Titanium}},\ }\href {https://doi.org/10.1103/PhysRevLett.123.105501} {\bibfield  {journal} {\bibinfo  {journal} {Physical Review Letters}\ }\textbf {\bibinfo {volume} {123}},\ \bibinfo {pages} {105501} (\bibinfo {year} {2019})}\BibitemShut {NoStop}%
\bibitem [{\citenamefont {Larikov}\ \emph {et~al.}(1967)\citenamefont {Larikov}, \citenamefont {Tyshkevich},\ and\ \citenamefont {Chernaya}}]{larikov_notitle_1967}%
  \BibitemOpen
  \bibfield  {author} {\bibinfo {author} {\bibfnamefont {L.~N.}\ \bibnamefont {Larikov}}, \bibinfo {author} {\bibfnamefont {V.~M.}\ \bibnamefont {Tyshkevich}},\ and\ \bibinfo {author} {\bibfnamefont {L.~F.}\ \bibnamefont {Chernaya}},\ }\href@noop {} {\bibfield  {journal} {\bibinfo  {journal} {Ukrainskyi fizychnyi zhurnal}\ }\textbf {\bibinfo {volume} {12}} (\bibinfo {year} {1967})}\BibitemShut {NoStop}%
\bibitem [{\citenamefont {Neumann}\ and\ \citenamefont {Tuijn}(2008)}]{neumann_self-diffusion_2008}%
  \BibitemOpen
  \bibfield  {author} {\bibinfo {author} {\bibfnamefont {G.}~\bibnamefont {Neumann}}\ and\ \bibinfo {author} {\bibfnamefont {C.}~\bibnamefont {Tuijn}},\ }\href {https://shop.elsevier.com/books/self-diffusion-and-impurity-diffusion-in-pure-metals/neumann/978-1-85617-511-1} {\emph {\bibinfo {title} {Self-diffusion and {Impurity} {Diffusion} in {Pure} {Metals}}}},\ \bibinfo {edition} {1st}\ ed.,\ Vol.~\bibinfo {volume} {14}\ (\bibinfo {year} {2008})\BibitemShut {NoStop}%
\bibitem [{\citenamefont {Erley}\ and\ \citenamefont {Wagner}(1974)}]{erley_diffusion_1974}%
  \BibitemOpen
  \bibfield  {author} {\bibinfo {author} {\bibfnamefont {W.}~\bibnamefont {Erley}}\ and\ \bibinfo {author} {\bibfnamefont {H.}~\bibnamefont {Wagner}},\ }\bibfield  {title} {\bibinfo {title} {The diffusion coefficients in the system molybdenum–tungsten},\ }\href {https://doi.org/10.1002/pssa.2210250212} {\bibfield  {journal} {\bibinfo  {journal} {physica status solidi (a)}\ }\textbf {\bibinfo {volume} {25}},\ \bibinfo {pages} {463} (\bibinfo {year} {1974})}\BibitemShut {NoStop}%
\bibitem [{\citenamefont {Henkelman}\ \emph {et~al.}(2000)\citenamefont {Henkelman}, \citenamefont {Uberuaga},\ and\ \citenamefont {Jónsson}}]{henkelman_climbing_2000}%
  \BibitemOpen
  \bibfield  {author} {\bibinfo {author} {\bibfnamefont {G.}~\bibnamefont {Henkelman}}, \bibinfo {author} {\bibfnamefont {B.~P.}\ \bibnamefont {Uberuaga}},\ and\ \bibinfo {author} {\bibfnamefont {H.}~\bibnamefont {Jónsson}},\ }\bibfield  {title} {\bibinfo {title} {A climbing image nudged elastic band method for finding saddle points and minimum energy paths},\ }\href {https://doi.org/10.1063/1.1329672} {\bibfield  {journal} {\bibinfo  {journal} {The Journal of Chemical Physics}\ }\textbf {\bibinfo {volume} {113}},\ \bibinfo {pages} {9901} (\bibinfo {year} {2000})}\BibitemShut {NoStop}%
\bibitem [{\citenamefont {Henkelman}\ and\ \citenamefont {Jónsson}(2000)}]{henkelman_improved_2000}%
  \BibitemOpen
  \bibfield  {author} {\bibinfo {author} {\bibfnamefont {G.}~\bibnamefont {Henkelman}}\ and\ \bibinfo {author} {\bibfnamefont {H.}~\bibnamefont {Jónsson}},\ }\bibfield  {title} {\bibinfo {title} {Improved tangent estimate in the nudged elastic band method for finding minimum energy paths and saddle points},\ }\href {https://doi.org/10.1063/1.1323224} {\bibfield  {journal} {\bibinfo  {journal} {The Journal of Chemical Physics}\ }\textbf {\bibinfo {volume} {113}},\ \bibinfo {pages} {9978} (\bibinfo {year} {2000})}\BibitemShut {NoStop}%
\bibitem [{\citenamefont {Perdew}\ \emph {et~al.}(1996)\citenamefont {Perdew}, \citenamefont {Burke},\ and\ \citenamefont {Ernzerhof}}]{perdew_generalized_1996}%
  \BibitemOpen
  \bibfield  {author} {\bibinfo {author} {\bibfnamefont {J.~P.}\ \bibnamefont {Perdew}}, \bibinfo {author} {\bibfnamefont {K.}~\bibnamefont {Burke}},\ and\ \bibinfo {author} {\bibfnamefont {M.}~\bibnamefont {Ernzerhof}},\ }\bibfield  {title} {\bibinfo {title} {Generalized {Gradient} {Approximation} {Made} {Simple}},\ }\href {https://doi.org/10.1103/PhysRevLett.77.3865} {\bibfield  {journal} {\bibinfo  {journal} {Physical Review Letters}\ }\textbf {\bibinfo {volume} {77}},\ \bibinfo {pages} {3865} (\bibinfo {year} {1996})}\BibitemShut {NoStop}%
\bibitem [{\citenamefont {Kresse}\ and\ \citenamefont {Furthmüller}(1996)}]{kresse_efficiency_1996}%
  \BibitemOpen
  \bibfield  {author} {\bibinfo {author} {\bibfnamefont {G.}~\bibnamefont {Kresse}}\ and\ \bibinfo {author} {\bibfnamefont {J.}~\bibnamefont {Furthmüller}},\ }\bibfield  {title} {\bibinfo {title} {Efficiency of ab-initio total energy calculations for metals and semiconductors using a plane-wave basis set},\ }\href {https://doi.org/10.1016/0927-0256(96)00008-0} {\bibfield  {journal} {\bibinfo  {journal} {Computational Materials Science}\ }\textbf {\bibinfo {volume} {6}},\ \bibinfo {pages} {15} (\bibinfo {year} {1996})}\BibitemShut {NoStop}%
\bibitem [{\citenamefont {Monkhorst}\ and\ \citenamefont {Pack}(1976)}]{monkhorst_special_1976}%
  \BibitemOpen
  \bibfield  {author} {\bibinfo {author} {\bibfnamefont {H.~J.}\ \bibnamefont {Monkhorst}}\ and\ \bibinfo {author} {\bibfnamefont {J.~D.}\ \bibnamefont {Pack}},\ }\bibfield  {title} {\bibinfo {title} {Special points for {Brillouin}-zone integrations},\ }\href {https://doi.org/10.1103/PhysRevB.13.5188} {\bibfield  {journal} {\bibinfo  {journal} {Physical Review B}\ }\textbf {\bibinfo {volume} {13}},\ \bibinfo {pages} {5188} (\bibinfo {year} {1976})}\BibitemShut {NoStop}%
\end{thebibliography}%
\end{document}


\title{Supplementary Information:\\Diffusion coefficients of multi-principal element alloys from first principles}
\author{Damien K. J. Lee}
\affiliation{Laboratory of materials design and simulation (MADES), Institute of Materials, \'{E}cole Polytechnique F\'{e}d\'{e}rale de Lausanne}
\affiliation{National Centre for Computational Design and Discovery of Novel Materials (MARVEL), \'{E}cole Polytechnique F\'{e}d\'{e}rale de Lausanne}
\author{Anirudh Raju Natarajan}
\affiliation{Laboratory of materials design and simulation (MADES), Institute of Materials, \'{E}cole Polytechnique F\'{e}d\'{e}rale de Lausanne}
\affiliation{National Centre for Computational Design and Discovery of Novel Materials (MARVEL), \'{E}cole Polytechnique F\'{e}d\'{e}rale de Lausanne}

\maketitle

\section{Decomposition of the site basis functions for Local Cluster Expansions}
Factorizing the full expression of the LCE yields:
\label{sec:LCE_decomposition}
\begin{align}
    \Delta E_{KRA}(\vec{\sigma}_L, \sigma_H)
    &=K_0+K_{\gamma}\psi(\sigma_H)
    +\sum_{\alpha\in L} K_{\alpha}\Phi_{\alpha}(\vec{\sigma}_L)
    +\sum_{\alpha\in L,\beta\in H} K_{\alpha\beta}\Phi_{\alpha}(\vec{\sigma}_L)\psi(\sigma_H)
\label{eqn:cluster_expansion_with_hop}
\end{align}

It can be shown that using the occupational basis function for the hopping species is functionally equivalent to constructing separate models for each hopping species, which has been the approach adopted in previous studies. For example, consider a binary alloy composed of species A and B, with a single non-constant site basis function $\psi$ defined as $\psi = 0$ for species A and $\psi = 1$ for species B. If species A is the migrating atom, then $\psi(\sigma_H) = 0$. Substituting this into \cref{eqn:cluster_expansion_with_hop} eliminates the species-dependent term, reducing the equation to:
\begin{equation}
\label{eqn:cluster_expansion_with_A}
\Delta E_{KRA}(\vec{\sigma}_L,0) = K_0 + \sum_{\alpha\in L} K_{\alpha} \Phi_{\alpha}(\vec{\sigma}_L)
\end{equation}

This is identical to a local cluster expansion trained solely on migration events involving species A. Now, consider the case where species B is the migrating atom, so that $\psi(\sigma_H) = 1$. The cluster expansion then becomes:
\begin{equation}
\label{eqn:cluster_expansion_with_B}
    \Delta E_{KRA}(\vec{\sigma}_L,1) = 
    (K_0 + K\gamma) 
    +\sum_{\alpha} (K_{\alpha} + K_{\alpha\beta}) \Phi_{\alpha}(\vec{\sigma}_L)
\end{equation}

This expression differs from \cref{eqn:cluster_expansion_with_A} only in the values of the expansion coefficients. In effect, it represents a separate but structurally identical model, with the expansion coefficients shifted by species-specific parameters $K_\gamma$ and $K_{\alpha\beta}$. Including the hopping species as a variable in the model therefore allows a unified treatment of all migrating species within a single formalism. This avoids the need to construct and train multiple separate models, while still allowing the model to differentiate between species-specific migration behavior \cite{zhang_kinetically_2019}.

\section{Symmetry equivalence in Local Cluster Expansions} \label{sec:symmetry}
Symmetry reduces the number of terms in the local cluster expansion. These symmetry operations are obtained from the undecorated lattice of the parent crystal. As an illustrative example, consider the following LCE model for the 2D square lattice in \cref{main:fig:kra} involving only point clusters with symmetrically equivalent clusters shown in the same color:
\begin{figure}[h!]
    \centering
    \includegraphics[width=0.4\linewidth]{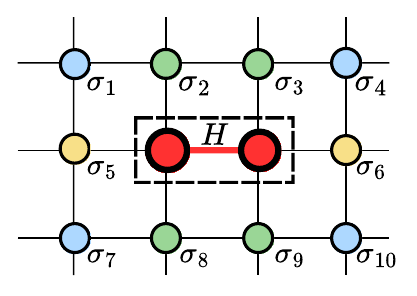}
    \caption{2D square lattice with symmetry equivalent point clusters grouped according to the same color.}
    \label{fig:symmetry_1}
\end{figure}

\begin{figure}[h!]
    \centering
    \includegraphics[width=1\linewidth]{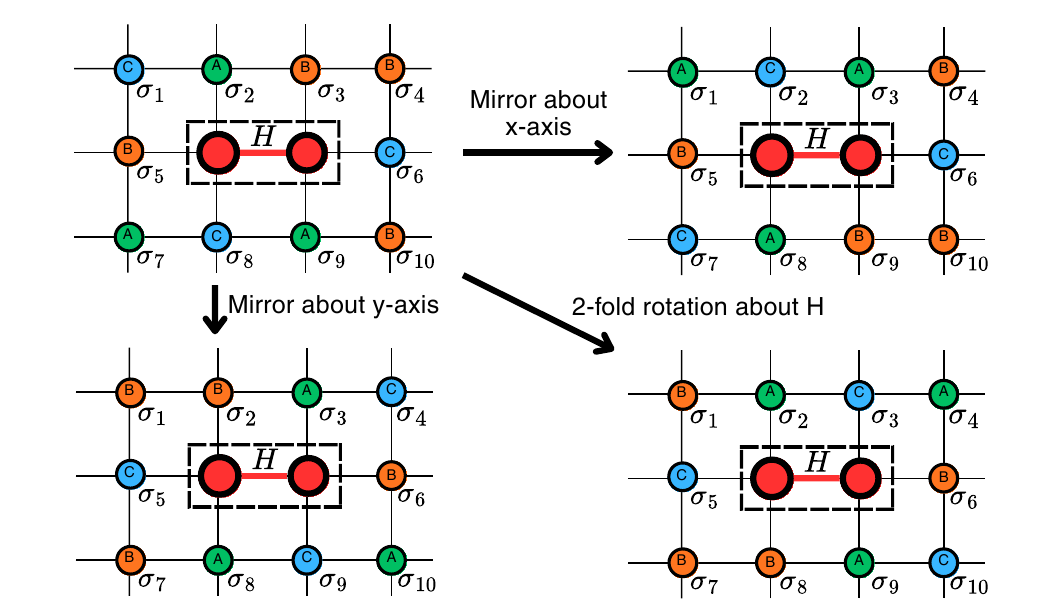}
    \caption{Symmetry operations on the 2D square lattice which leave the KRA unchanged.}
    \label{fig:symmetry_2}
\end{figure}

Only seven independent coefficients ($V_0$ to $V_6$) are required to evaluate the KRA:
\begin{equation}
\begin{split}
\Delta E_{\textrm{KRA}} = V_0
&+ V_1[\phi_{1}(\sigma_1)+\phi_{1}(\sigma_4)+\phi_{1}(\sigma_7)+\phi_{1}(\sigma_{10})] \\
&+ V_2[\phi_{1}(\sigma_2)+\phi_{1}(\sigma_3)+\phi_{1}(\sigma_8)+\phi_{1}(\sigma_{9})] + V_3[\phi_{1}(\sigma_5)+\phi_{1}(\sigma_6)] \\
&+ V_4[\phi_{2}(\sigma_1)+\phi_{2}(\sigma_4)+\phi_{2}(\sigma_7)+\phi_{2}(\sigma_{10})] \\
&+ V_5[\phi_{2}(\sigma_2)+\phi_{2}(\sigma_3)+\phi_{2}(\sigma_8)+\phi_{2}(\sigma_{9})] + V_6[\phi_{2}(\sigma_5)+\phi_{2}(\sigma_6)]
\end{split}
\end{equation}
where each site may be occupied by species A, B, or C, which requires two site basis functions, denoted $\phi_{1}$ and $\phi_{2}$, in addition to the constant site basis function. The symmetry-related orderings shown in \cref{fig:symmetry_2} must therefore yield the same KRA value.

\begin{figure}[h!]
    \centering
    \includegraphics[width=1\linewidth]{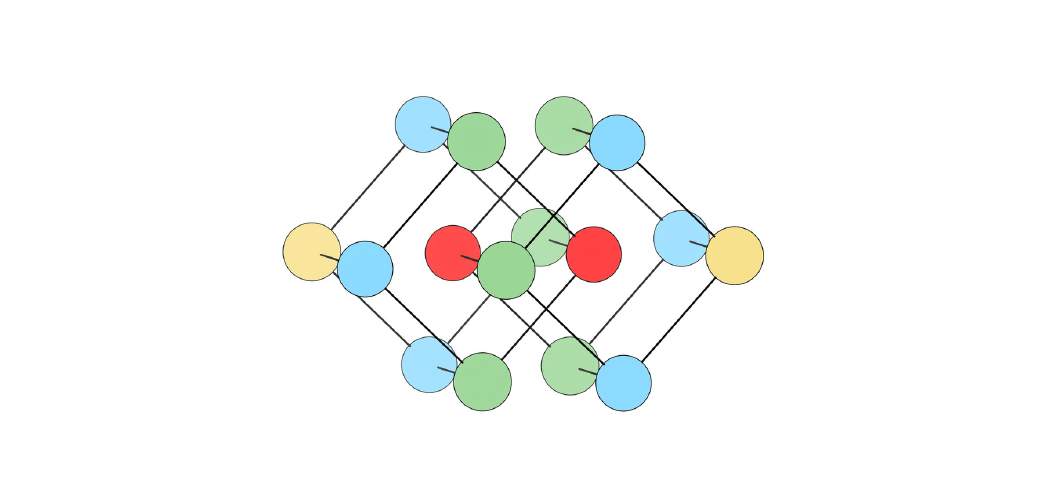}
    \caption{Symmetry equivalent point clusters grouped by the same color in the bcc lattice.}
    \label{fig:bcc_symmetry}
\end{figure}

Grouping these equivalent clusters to share expansion coefficients reduces the number of independent terms and, by construction, forces the model to predict identical KRA values for symmetry-related orderings. This grouping is a property of the lattice geometry, not of any particular decoration, so it holds for arbitrary chemical arrangements, including the random alloy.
For the bcc lattice used in this work, the operations are drawn from the corresponding subgroup of the bcc space group. The symmetrically equivalent point clusters among the nearest neighbors of the hop cluster (red) in the undecorated crystal are shown in \cref{fig:bcc_symmetry}.

\section{Parity plot of DFT and MACE-DFT calculations}
\label{sec:mace_dft_comparison}
\begin{figure}[hbt!]
    \centering
    \includegraphics[width=0.45\linewidth]{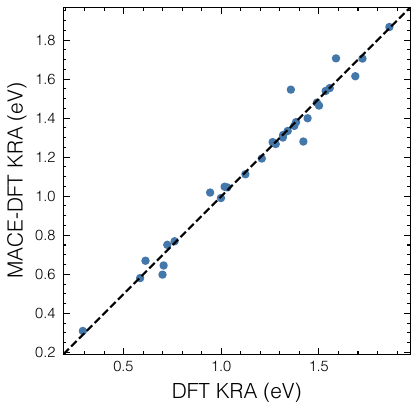}
    \caption{Parity plot of KRA values using the MACE-DFT workflow and full DFT calculations. The MAE achieved here is 36.6 meV.}
    \label{fig:parity_plot_mace_dft}
\end{figure}

\newpage
\section{Parity plot of eLCE predictions vs MACE-DFT calculations}
\label{sec:parity_plot}
\begin{figure}[hbt!]
    \centering
    \includegraphics[width=0.45\linewidth]{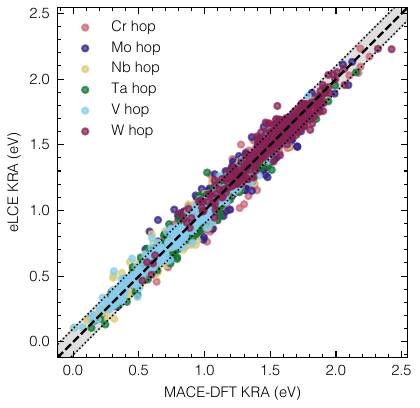}
    \caption{Parity plot of KRA values predicted from the 2-3-eLCE model and MACE-DFT calculations. The test set MAE here is 69.9 meV.}
    \label{fig:parity_plot_eLCE}
\end{figure}

\section{Parity plot of DFT and eCE formation energy predictions}
\label{sec:parity_plot_energy}
\begin{figure}[hbt!]
    \centering
    \includegraphics[width=0.45\linewidth]{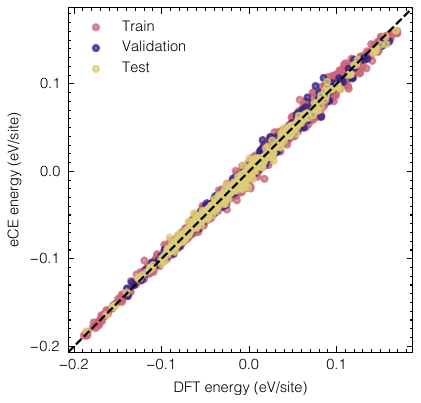}
    \caption{Parity plot of formation energies calculated from DFT and predicted from the 3-eCE model. The test set MAE here is 3.5 meV/site.}
    \label{fig:parity_plot_energy}
\end{figure}

\section{Analytical expressions for kinetic quantities}
\label{sec:equations}
\subsection{Onsager transport coefficients}
\citeauthor{moleko_self-consistent_1989} \cite{moleko_self-consistent_1989} derived analytical expressions for a thermodynamically ideal and kinetically simple binary alloy:

\begin{align}
    & \tilde{L}_{A A}=x_{Va} x_{A} \rho a^{2} \Gamma_{A}\left(1-\frac{2 x_{B} \Gamma_{A}}{\Lambda}\right), \\
    & \tilde{L}_{B B}=x_{Va} x_{B} \rho a^{2} \Gamma_{B}\left(1-\frac{2 x_{A} \Gamma_{B}}{\Lambda}\right), \\
    & \tilde{L}_{A B}=\frac{2 \rho a^{2} \Gamma_{A} \Gamma_{B} x_{Va} x_{A} x_{B}}{\Lambda} 
\end{align}
with
\begin{align}
    \Lambda
    &= \frac12 (F+2)(x_A \Gamma_A + x_B \Gamma_B)
     - \Gamma_A - \Gamma_B + 2(x_A \Gamma_B + x_B \Gamma_A) \nonumber \\ 
    & + \sqrt{\left(\frac12 (F+2)(x_A \Gamma_A + x_B \Gamma_B)
     - \Gamma_A - \Gamma_B
    \right)^{2} + 2 F \Gamma_A \Gamma_B } 
\end{align}
and
\begin{equation}
    F = \frac{f_0}{1-f_0}
\end{equation}
$a$ is the hop distance and $\rho=\frac{Z}{2d}$ where $Z$ is the coordination number and $d$ is the dimensionality of the crystal. $f_0$ is the correlation factor of the Bravais lattice (0.727 for bcc).

\subsection{Thermodynamic Factor}
The free energy of a thermodynamically ideal alloy is \cite{van_der_ven_vacancy_2010}:
\begin{equation}
    g(x_A,x_B)= x_{Va}\Delta g_{Va} + k_BT[x_A \ln x_A + x_B \ln x_B + x_{Va} \ln x_{Va}]
\end{equation}
Differentiation of the above yields:
\begin{equation}
    \mathbf{\tilde{\Theta}} = 
    \begin{pmatrix}
    \frac{1 - x_B}{x_A x_{Va}} & \frac{1}{x_{Va}} \\
    \frac{1}{x_{Va}} & \frac{1 - x_A}{x_B x_{Va}}
    \end{pmatrix}
\end{equation}

\subsection{Tracer diffusion coefficients}
The tracer diffusion coefficients of pure elements have the following expression:
\begin{align}
    &D^*_{Va} = D_0 \exp{\left(-\frac{\Delta E}{k_B T} \right)} \\
    &D^*_{i} = x_{Va}f_0 D_0 \exp{\left(-\frac{\Delta E}{k_B T} \right)}
\end{align}
where 
\begin{equation}
    D_0 = a^2 \rho \nu^*
\end{equation}

For alloys, the vacancy tracer diffusion coefficient can be obtained from the largest eigenvalue of $\mathbf{D}$.

\subsection{Correlation factors}
\citeauthor{manning_correlation_1971} \cite{manning_correlation_1971} derived analytical equations for the correlation factor in a thermodynamically ideal and kinetically simple alloy as follows \cite{mishin_monte_1997}:
\begin{align}
    &f_{i\neq Va}= \frac{H}{2\Gamma_i+H} \\
    & H = h+ \left(h^2+\frac{4\Gamma_A\Gamma_B f_0}{1-f_0} \right)^{0.5} \\
    & h = \frac{x_A\Gamma_A+x_B\Gamma_B}{1-f_0}-(\Gamma_A + \Gamma_B) 
\end{align}
The vacancy correlation factor is as follows:
\begin{equation}
    f_{Va}=\frac{f_Ax_A\Gamma_A+f_Bx_B\Gamma_B}{f_0(x_A\Gamma_A+x_B\Gamma_B)}
\end{equation}

\section{Correlation between KRA spectrum and lattice mismatch}
\label{sec:lattice_mismatch}
\begin{figure}[hbt!]
    \centering
    \includegraphics[width=0.5\linewidth]{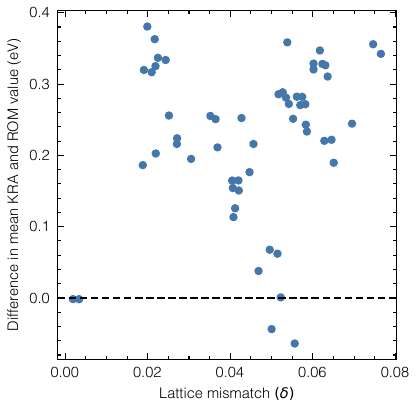}
    \caption{Scatter plot of difference between mean of KRA spectrum and ROM value of various equiatomic alloys as a function of lattice mismatch defined in Ref. \cite{daw_sluggish_2021}.}
    \label{fig:lattice_mismatch}
\end{figure}

\section{Comparison with experimental data and benchmarks on modeling assumptions} \label{sec:benchmarks}

\subsection{Experimental measurements}
\begin{figure}[hbt!]
    \centering
    \includegraphics{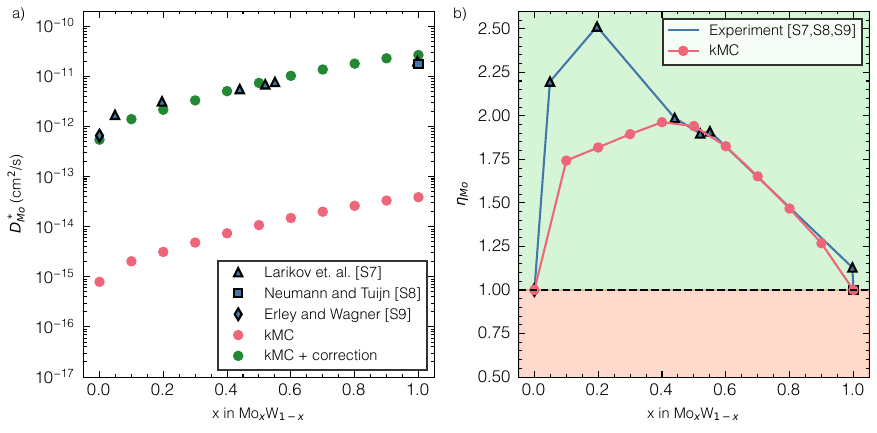}
    \caption{\textbf{a)} Tracer diffusion coefficients of Mo in the Mo--W alloy at 2173~K, showing experimental measurements \cite{larikov_notitle_1967,neumann_self-diffusion_2008,erley_diffusion_1974}, kMC results from this work, and the same kMC results after applying an empirical, composition-independent correction. \textbf{b)} Enhancement factor of Mo diffusion, $\eta_{Mo}$, obtained from the experimental and kMC tracer coefficients in panel a. No correction enters $\eta_{Mo}$, which is defined relative to the rule-of-mixtures estimate.}
    \label{fig:experimental}
\end{figure}

To assess the quantitative accuracy of the model, we searched the experimental literature for measurements of non-dilute transport coefficients in the systems studied here. Such measurements are scarce. Of the systems studied here, the Mo--W alloy provides data suitable for benchmarking the predictions of the present framework against experiment. \Cref{fig:experimental}a shows the tracer diffusion coefficient $D^*_{Mo}$ across the entire composition range as measured by \citeauthor{larikov_notitle_1967} \cite{larikov_notitle_1967} at 2173~K, together with values for Mo self-diffusion in pure Mo and W \cite{neumann_self-diffusion_2008,erley_diffusion_1974}. We use the Mo enhancement factor $\eta_{Mo}$ for the comparison, because the vacancy has no radiotracer analogue and $\eta_{Va}$ is therefore not directly measurable.

The comparison of the enhancement factor for Mo in Mo--W is shown in \cref{fig:experimental}b. Predicted and measured enhancement factors agree to within a few percent for $x_{Mo} \gtrsim 0.4$, and at $x_{Mo} \approx 0.4$ model and experiment both give $\eta_{Mo} \approx 2.0$. The largest discrepancy, approximately 25\%, occurs near $x_{Mo} \approx 0.2$, where the simulations underestimate the measured enhancement. Both prediction and experiment place $\eta_{Mo}$ above unity at all intermediate compositions, so the two agree that Mo diffusion in Mo--W is anti-sluggish throughout. Contributions from grain-boundary and other short-circuit diffusion paths, which our model does not represent, may account for part of the remaining discrepancy, as may the residual error of the eLCE model itself. Obtained without empirical correction, this level of agreement indicates that the model captures the physics of non-dilute diffusion in these alloys.

The comparison of absolute tracer diffusivities involves an empirical, composition-independent correction. This correction is applied only to assess the absolute magnitude of the calculated diffusivities and does not enter our analysis of sluggish and anti-sluggish diffusion. \Cref{fig:experimental}a compares the absolute Mo tracer diffusion coefficient from experiment, from our kMC simulations directly, and after this correction. Unlike the enhancement factor, the absolute diffusivities carry a substantial and well-understood systematic offset. As is common in first-principles diffusion modeling \cite{wu_high-throughput_2016,zou_impurity_2019,naghavi_diffusion_2017,shang_comprehensive_2016}, the absolute $D^*$ values are sensitive to the neglected vibrational contributions to the vacancy formation free energy, of order 0.5~eV in these refractory elements \cite{forslund_thermodynamic_2023}, and to the vibrational prefactor entering the transition rates. Because these contributions are largely systematic across compositions, they cancel in the enhancement factor, which is why $\eta_{Mo}$ agrees with experiment while the absolute $D^*$ values are offset. When absolute values are required, they are recoverable by setting the prefactor from the pure-element and dilute-limit diffusivities, following the correction proposed by \citeauthor{wu_high-throughput_2016} \cite{wu_high-throughput_2016}. \citeauthor{wu_high-throughput_2016} demonstrated that even tracer diffusion coefficients of pure elements computed using DFT deviate significantly from experimental values, and proposed a correction of the form
\begin{equation}
D_{\textrm{corrected}}^* = 
D_{\textrm{computed}}^* \times 
A_{\textrm{shift}}\exp\left( -\frac{E_{\textrm{shift}}}{k_B T} \right)
\end{equation}
where $A_{\textrm{shift}}$ and $E_{\textrm{shift}}$ collect distinct sources of systematic error. We set $A_{\textrm{shift}}=\exp(\Delta S_{Va}/k_B)$, with $T\Delta S_{Va}=0.5$~eV estimated from previous studies \cite{wang_nonlinear_2022,forslund_thermodynamic_2023,lapointe_anomalous_2025}, so that this term captures the vibrational contribution to the vacancy formation free energy that the model omits and that we identify as a limitation in the Discussion of the main text. A further correction of $E_{\textrm{shift}}=-0.45$~eV, determined empirically, brings the predictions into agreement with experiment. This residual term collects the remaining systematic errors, including inaccuracies in the DFT energies \cite{forslund_thermodynamic_2023} and fitting error of the cluster expansion \cite{lee_modeling_2026}. After applying this correction, the computed tracer diffusion coefficients agree with the measured values across the full composition range (\cref{fig:experimental}a).

As noted above, this correction does not enter our analysis, because the enhancement factors are defined relative to the corresponding rule-of-mixtures estimate, such that systematic errors common to the alloy and its pure-element references largely cancel. Furthermore, $T\Delta S_{Va}$ represents only one contribution to the prefactor correction $A_{\textrm{shift}}$, which may also contain other systematic uncertainties in first-principles calculations. The present level of agreement is consistent with the state of the art in first-principles diffusion modeling, and the metric underlying our conclusions, the enhancement factor, is in quantitative agreement with experiment.

\subsection{Benchmarks of modeling assumptions} 
\begin{figure}
    \centering
    \includegraphics[width=0.5\linewidth]{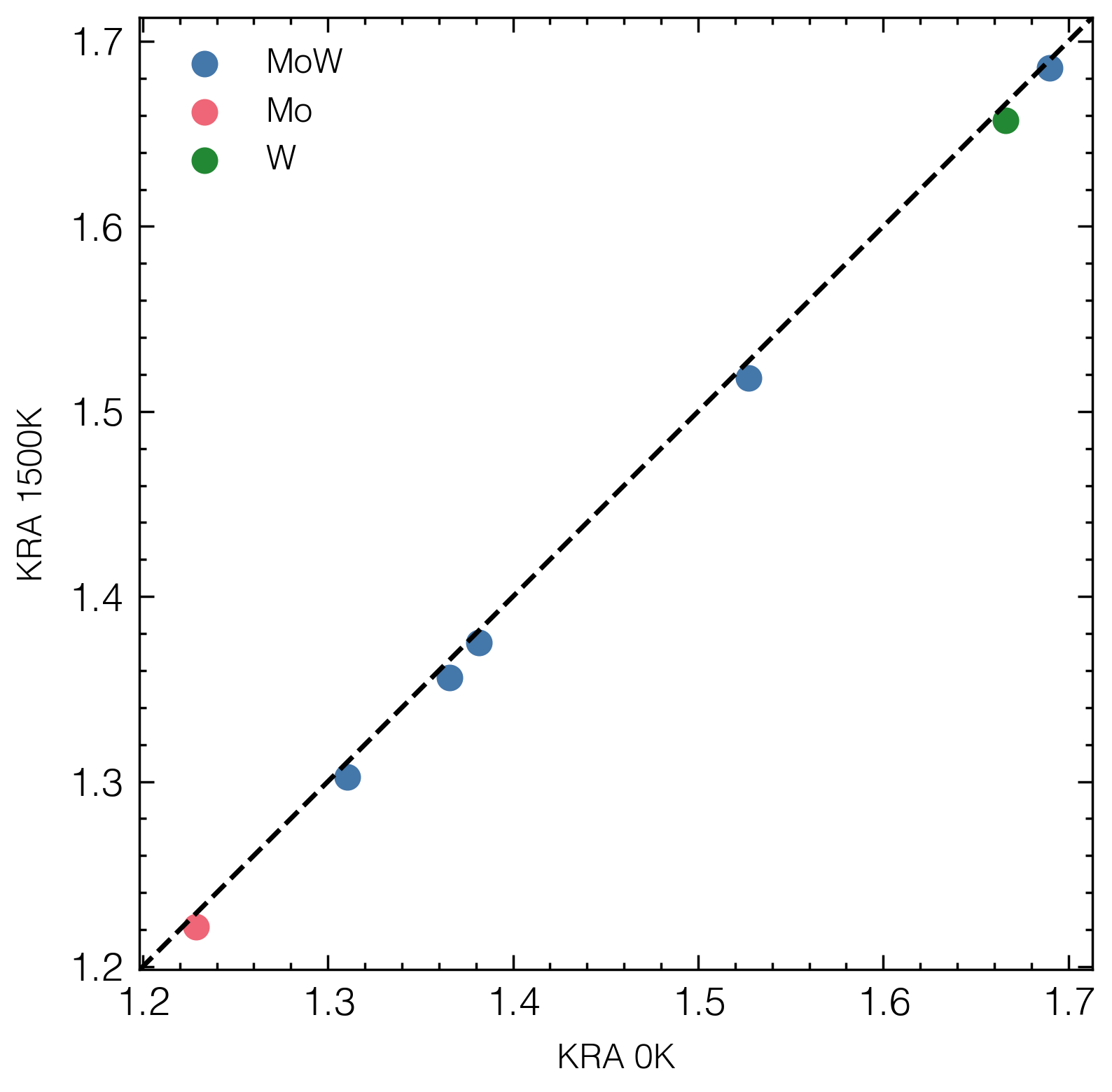}
    \caption{Parity plot of DFT-computed KRA values evaluated at the thermally expanded lattice parameters at 1500~K against those evaluated at the 0~K lattice parameters. Configurations are sampled across the KRA distribution of the Mo--W alloy (MoW), together with pure Mo and pure W. The dashed line marks parity. All values fall on or slightly below parity, corresponding to an average decrease of approximately 7~meV.}
    \label{fig:thermal_expansion}
\end{figure}

\textbf{Thermal expansion.} In the Mo--W alloy, thermal lattice expansion lowers migration barriers by reducing steric hindrance for the diffusing atom. To estimate the magnitude of this effect, we used experimental thermal expansion data for these alloys \cite{lu_theoretical_2005} and recomputed migration barriers for configurations sampled uniformly across the KRA distribution of Mo--W. As shown in \cref{fig:thermal_expansion}, the barriers decrease on average by approximately 7~meV at 1500~K. Propagating this shift through the enhancement factor,
\begin{equation}
\eta_{Va}(1500\,\mathrm{K})=\eta_{Va}(0\,\mathrm{K})\frac{
\exp\left(
\frac{-\Delta E_{\textrm{expansion}}}
{k_B T} \right)
_{\textrm{alloy}}}
{{\langle\exp\left( \frac{-\Delta E_{\textrm{expansion}}}
{k_B T}\right)_{\textrm{pure}}\rangle}}
\end{equation}
yields a correction of approximately 1.005, a 0.5\% change in $\eta_{Va}$. This is negligible relative to the larger variations in $\eta_{Va}$ discussed in this work and does not affect our conclusions regarding sluggish diffusion.

\textbf{Vibrational contributions.} Vibrational effects modify both vacancy formation free energies and migration barriers. Determining harmonic and anharmonic vibrational contributions accurately is challenging even for pure elements. Near-\emph{ab initio} accuracy has been achieved for pure refractory metals only through combined DFT and machine-learning approaches \cite{zhang_ab_2025,forslund_thermodynamic_2023,jung_high-accuracy_2023}, and extending these to concentrated multicomponent alloys is currently intractable. As discussed above, the dominant consequence for our results is a systematic shift in the absolute tracer diffusion coefficients, which largely cancels in the enhancement factor. Our benchmark confirms that the compositional trends and the predicted anti-sluggish behavior across the full Mo--W composition range are preserved (\cref{fig:experimental}b).
The absolute magnitude of $D^*_{Mo}$ is more sensitive to the vibrational treatment, primarily through uncertainties in the attempt frequency. As shown in \cref{fig:experimental}a, the absolute tracer coefficients are nonetheless brought into close agreement with experiment by the empirical, composition-independent correction described above, consistent with prior work \cite{wu_high-throughput_2016,zou_impurity_2019}. Absolute diffusion coefficients are therefore also accessible within the present framework, while the enhancement factors central to our conclusions remain parameter-free.

\textbf{Magnetic effects.} Magnetic effects are expected to play a limited role here. The alloys are studied at high temperature in their disordered phase, where long-range magnetic order is substantially weakened, and most compositions in the V--Cr--Nb--Mo--Ta--W space are dominated by non-magnetic refractory elements. Magnetic disorder can quantitatively modify vacancy formation and migration energies, as shown for Fe--Cr and Ni--Cr alloys \cite{senninger_atomistic_2014,li_magnetochemical_2022}. Chromium and vanadium are the magnetic constituents in our composition space, and we accordingly note that predictions for Cr-rich and V-rich compositions carry additional uncertainty from this source. Coupling magnetic and chemical degrees of freedom within the eCE/eLCE framework is feasible and represents a clear direction for future work, particularly for systems in which magnetic ordering persists at the temperatures of interest.

\bibliography{references}